%% file: main.tex
\documentclass[sigconf]{acmart}

\AtBeginDocument{
  }

\copyrightyear{2026}
\acmYear{2026}
\setcopyright{cc}
\setcctype{by-nc-nd}
\acmConference[ICSE-SEIP '26]{2026 IEEE/ACM 48th International Conference on Software Engineering}{April 12--18, 2026}{Rio de Janeiro, Brazil}
\acmBooktitle{2026 IEEE/ACM 48th International Conference on Software Engineering (ICSE-SEIP '26), April 12--18, 2026, Rio de Janeiro, Brazil}
\acmPrice{}
\acmDOI{10.1145/3786583.3786908}
\acmISBN{979-8-4007-2426-8/2026/04}
\usepackage{multirow}   
\usepackage[utf8]{inputenc}
\usepackage{colortbl}
\usepackage{enumitem}
\usepackage{subcaption}
\usepackage{cleveref}
\usepackage{url}
\usepackage{ifthen}
\usepackage{soul} 
\usepackage{algorithm}
\usepackage{algorithmic}
\usepackage{listings}
\usepackage{framed}
\definecolor{analysisbg}{RGB}{238,242,255}   
\definecolor{analysisbar}{RGB}{17,24,39} 

\makeatletter
\newenvironment{paperbox}[2][]{
  \def\pb@bg{#2}
  \def\FrameCommand##1{
    {\setlength{\fboxsep}{\FrameSep}
     \fcolorbox{black!18}{\pb@bg}{##1}}
  }
  \MakeFramed{\advance\hsize-\width \FrameRestore}
  \noindent
  \if\relax\detokenize{#1}\relax\else
    {\setlength{\fboxsep}{4pt}
     \colorbox{analysisbar}{
       \parbox{\dimexpr\linewidth-2\fboxsep\relax}{\color{white}\bfseries #1}
     }}
    \par\smallskip
  \fi
  \small
}{
  \endMakeFramed
}
\makeatother

\definecolor{algbg}{RGB}{248,250,252}
\definecolor{algbar}{RGB}{0,0,0}

\makeatletter
\newenvironment{Algobox}[1]{
  \def\FrameCommand##1{
    {\setlength{\fboxsep}{\FrameSep}
     \fcolorbox{black}{algbg}{##1}}
  }
  \MakeFramed{\advance\hsize-\width \FrameRestore}
  \noindent
  {\setlength{\fboxsep}{4pt}
   \colorbox{algbar}{
     \parbox{\dimexpr\linewidth-2\fboxsep\relax}{\color{white}\bfseries #1}
   }}
  \par\smallskip
}{
  \endMakeFramed
}
\makeatother

\definecolor{framedBG}{RGB}{235, 238, 245}      
\definecolor{framedBorder}{RGB}{17, 24, 39}   

\makeatletter
\renewenvironment{framed}{
  \def\FrameCommand##1{
    {\setlength{\fboxsep}{\FrameSep}
     \fcolorbox{framedBorder}{framedBG}{##1}}
  }
  \MakeFramed{\advance\hsize-\width \FrameRestore}
  \noindent
}{
  \endMakeFramed
}
\makeatother

\definecolor{vscPink}{HTML}{C586C0}   % imports: from, import, as
\definecolor{vscYellow}{HTML}{DCDCAA} % def
\definecolor{vscOrange}{HTML}{CE9178} % comments
\definecolor{revyellow}{RGB}{255,249,196} % warm pastel yellow
\sethlcolor{revyellow}

\lstdefinestyle{pyintro}{
  language=Python,
  basicstyle=\ttfamily\footnotesize, 
  columns=fullflexible,
  keepspaces=true,
  showstringspaces=false,
  breaklines=true,
  breakatwhitespace=true,
  commentstyle=\color{vscOrange},      % comments -> orange
  keywordstyle=\color{black},          % neutralize default keyword colors
  emph={def},                          % 'def' -> yellow
  emphstyle=\color{vscYellow},
  emph={[2]from,import,as},            % import keywords -> pink
  emphstyle={[2]\color{vscPink}},
}

\definecolor{problembg}{HTML}{E0EFFF}  % Light blue
\definecolor{goodalgobg}{HTML}{E0F8E0} % Light green
\definecolor{badalgobg}{HTML}{FFF0E0}  % Light orange
\definecolor{analysisbg}{HTML}{F5F5F5} % Light gray

\newboolean{showcomments}
\setboolean{showcomments}{false}
\ifthenelse{\boolean{showcomments}}
 { \newcommand{\mynote}[2]{
      \fbox{\bfseries\sffamily\scriptsize#1}
        {\small$\blacktriangleright$\textsf{\emph{#2}}$\blacktriangleleft$}}}
        { \newcommand{\mynote}[2]{}}

\newlength{\SlidingMaxColH}

\begin{document}

\title{Correctness isn’t Efficiency: Runtime Memory Divergence in LLM-Generated Code}

\author{Prateek Rajput$^\star$}\thanks{$^\star$ Prateek's research is in collaboration with Zortify}
\orcid{0000-0001-2345-6789}
\affiliation{
  \institution{University of Luxembourg}
  \city{Esch-sur-Alzette}
  \country{Luxembourg}
}
\email{prateek.rajput@uni.lu}

\author{Yewei Song$^\dagger$}\thanks{$^\dagger$ Yewei's research is in collaboration with BGL BNP Paribas}
\orcid{0000-0002-6314-7515}
\affiliation{
  \institution{University of Luxembourg}
  \city{Esch-sur-Alzette}
  \country{Luxembourg}
}
\email{yewei.song@uni.lu}

\author{Abdoul Aziz Bonkoungou$^\ddagger$}\thanks{$^\ddagger$ Aziz's research is in collaboration with B Medical Systems}
\orcid{0009-0002-2361-485X}
\affiliation{
  \institution{University of Luxembourg}
  \city{Esch-sur-Alzette}
  \country{Luxembourg}
}
\email{abdoul.bonkoungou@uni.lu}

\author{Iyiola E. Olatunji}
\orcid{0000-0002-0391-9202}
\affiliation{
  \institution{University of Luxembourg}
  \city{Esch-sur-Alzette}
  \country{Luxembourg}
}
\email{emmanuel.olatunji@uni.lu}

\author{Abdoul Kader Kabore}
\orcid{0000-0002-3151-9433}
\affiliation{
  \institution{University of Luxembourg}
  \city{Esch-sur-Alzette}
  \country{Luxembourg}
}
\email{abdoulkader.kabore@uni.lu}

\author{Jacques Klein}
\orcid{0000-0003-4052-475X}
\affiliation{
  \institution{University of Luxembourg}
  \city{Esch-sur-Alzette}
  \country{Luxembourg}
}
\email{jacques.klein@uni.lu}

\author{Tegawendé F. Bissyandé}
\orcid{0000-0001-7270-9869}
\affiliation{
  \institution{University of Luxembourg}
  \city{Esch-sur-Alzette}
  \country{Luxembourg}
}
\email{tegewende.bissyande@uni.lu}

\renewcommand{\shortauthors}{Rajput et al.}

\begin{abstract}
LLMs can produce functionally correct programs, yet correctness alone does not guarantee reliability. Two programs passing the same tests can exhibit drastically different runtime behavior, creating hidden risks such as performance bottlenecks and memory leaks. Despite this, the runtime consistency of LLM-generated code remains largely unexplored.
In this work, we introduce a framework to systematically quantify \emph{execution-time memory stability} across multiple correct generations for the same task. We propose a novel solution-level metric, \textbf{DMPD (Dynamic Mean Pairwise Distance)}, which uses \textbf{Dynamic Time Warping} to compare the shapes of memory usage profiles. These profiles, which we term \textbf{Monotonic Peak Profiles (MPPs)}, are transformed to suppress transient noise, enabling robust comparison. By aggregating these scores, we derive a model-level \textbf{Model Instability Score (MIS)}. Across the BigOBench and CodeContests benchmarks, we find substantial runtime divergence among correct solutions, revealing that instability often increases with higher sampling temperatures even as pass@1 improves. We also uncover exploratory correlations between our stability metrics and established software-engineering indicators (e.g., Cognitive and Cyclomatic Complexity), suggesting a link between operational behavior and code maintainability. These findings enable stability-aware selection of passing candidates in CI/CD pipelines, reducing operational risk without sacrificing correctness. Artifacts are available at \url{https://github.com/pkrajput/memory_profiling}.
\end{abstract}

%    <concept>
%        <concept_id>10003752.10003809</concept_id>
%        <concept_desc>Theory of computation~Design and analysis of algorithms</concept_desc>
%        <concept_significance>500</concept_significance>
%        </concept>
%  </ccs2012>

% \ccsdesc[500]{Software engineering~Software testing and debugging}
% \ccsdesc[500]{Theory of computation~Design and analysis of algorithms}
%\keywords{Algorithmic Stability, Large Language Models, Code Generation, Computational Complexity, Non-Functional Properties, Empirical Software Engineering}

\maketitle

\input{sections/introduction}

\input{sections/research_questions}
\input{sections/approach}
\input{sections/experiments}

\input{sections/results}

\input{sections/discussion}
\input{sections/threats_to_validity}
\input{sections/relwork}
\input{sections/conclusion}

\balance
\bibliographystyle{ACM-Reference-Format}
\bibliography{main}

\appendix
\end{document}

%% file: sections/introduction.tex
\section{Introduction}

\begin{figure}[H]
  \centering
  \captionsetup[subfigure]{skip=2pt}

  \begin{subfigure}[t]{0.49\linewidth}
    \centering
    \includegraphics[page=1,width=\linewidth]{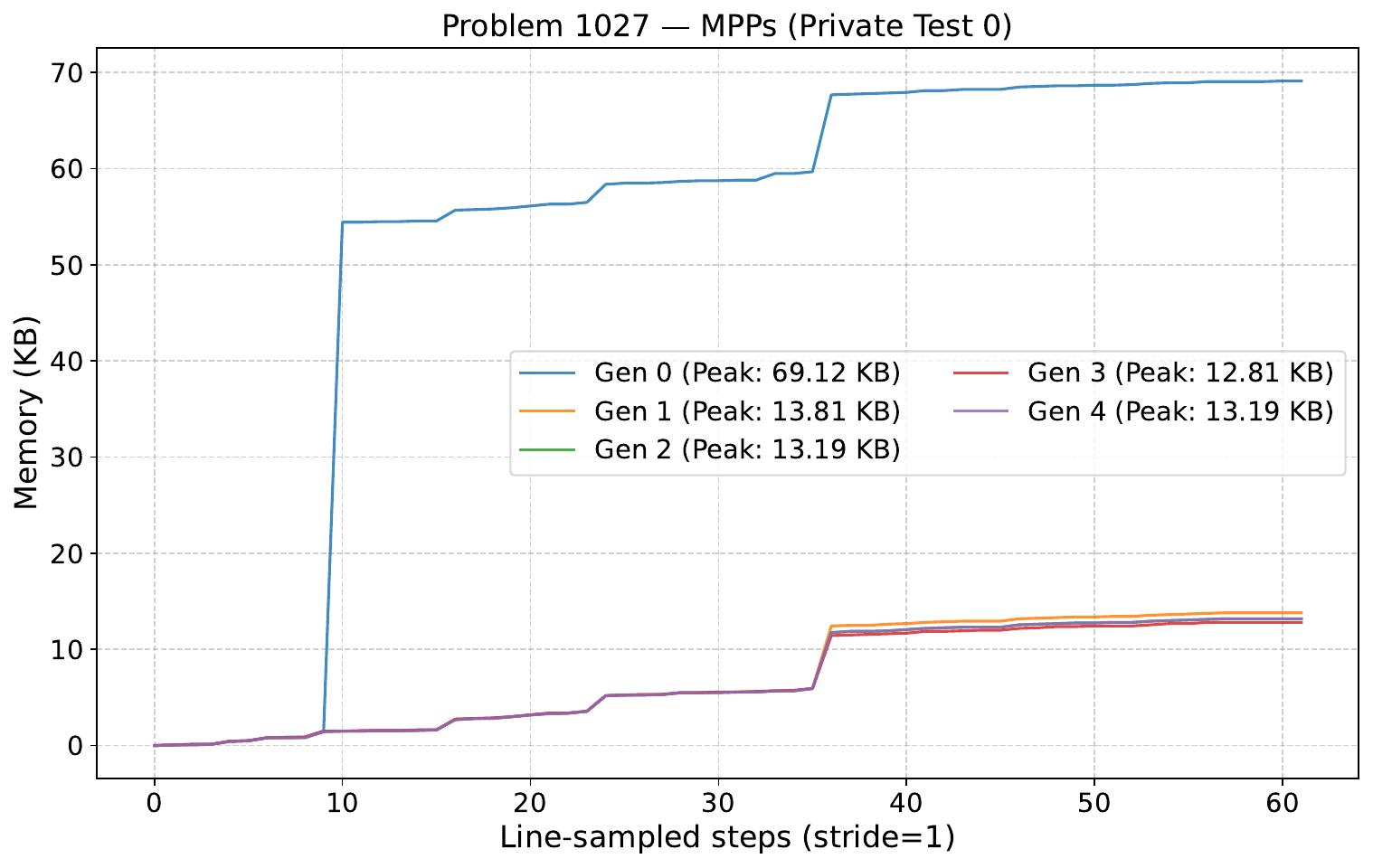}
    \phantomcaption\label{fig:mpp_divergence} 
  \end{subfigure}\hfill
  \begin{subfigure}[t]{0.49\linewidth}
    \centering
    \includegraphics[page=1,width=\linewidth]{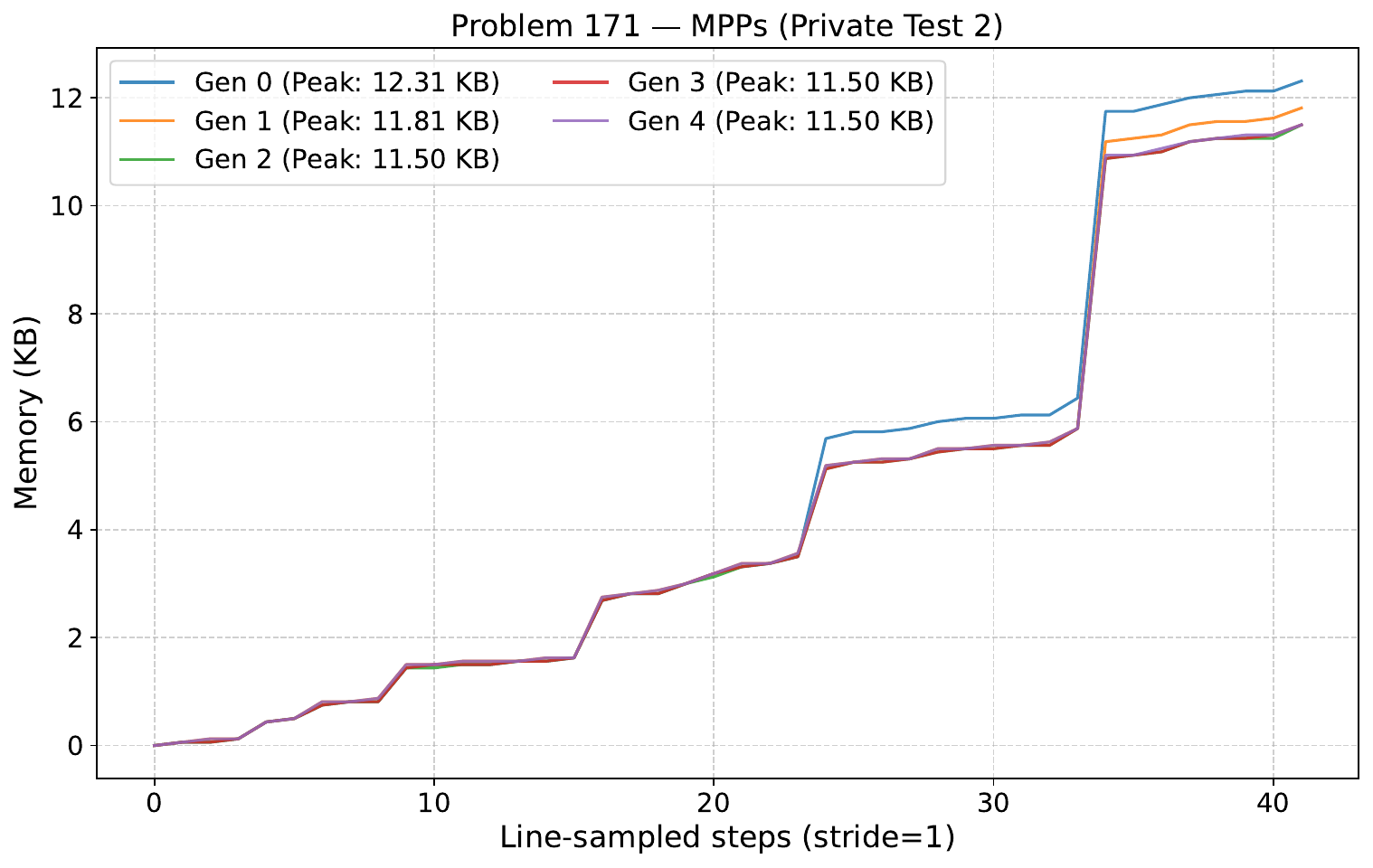}
    \phantomcaption\label{fig:mpp_consistency} 
  \end{subfigure}
  \vspace{-2.3em}
  \caption{Monotonic Peak Profiles (MPP) on \textsc{BigOBench} illustrating two regimes of runtime memory behavior across correct generations: left \emph{divergence} where one MPP differs significantly (high pairwise DMPD); right \emph{consistency} where profiles remain closely aligned (low pairwise DMPD).}
  \label{fig:mpp_profiles_execution_memory} 
\end{figure}

Large language models (LLMs) now routinely synthesize correct programs, with evaluation dominated by execution-based metrics such as pass@k on standardized suites \cite{chen2021evaluating,li2022competition,zhuo2024bigcodebench}. However, correctness alone does not characterize how a model \emph{behaves} at runtime once a solution has passed tests. In production, two equally correct solutions that exhibit different memory-allocation dynamics can have materially different cost and reliability profiles. Cloud platforms often charge in proportion to configured or consumed memory, and containerized deployments routinely fail due to out-of-memory (OOM) events when limits are exceeded. Therefore, beyond functional correctness, measuring \emph{runtime consistency} across a model’s multiple valid generations for the same problem is important.

Application-level memory traces are inherently temporal and noisy. In Python, reference counting and a cyclic garbage collector trigger non-deterministic reclamation and brief oscillations even for identical logic~\cite{pep454,tracemalloc,gcdocs}. System-level indicators (e.g., RSS) further blur the signal by mixing allocator policy, fragmentation, and non-Python allocations~\cite{valgrind2007}.

\begin{figure}[htbp]
  \centering
    \includegraphics[page=1,width=0.75\linewidth]{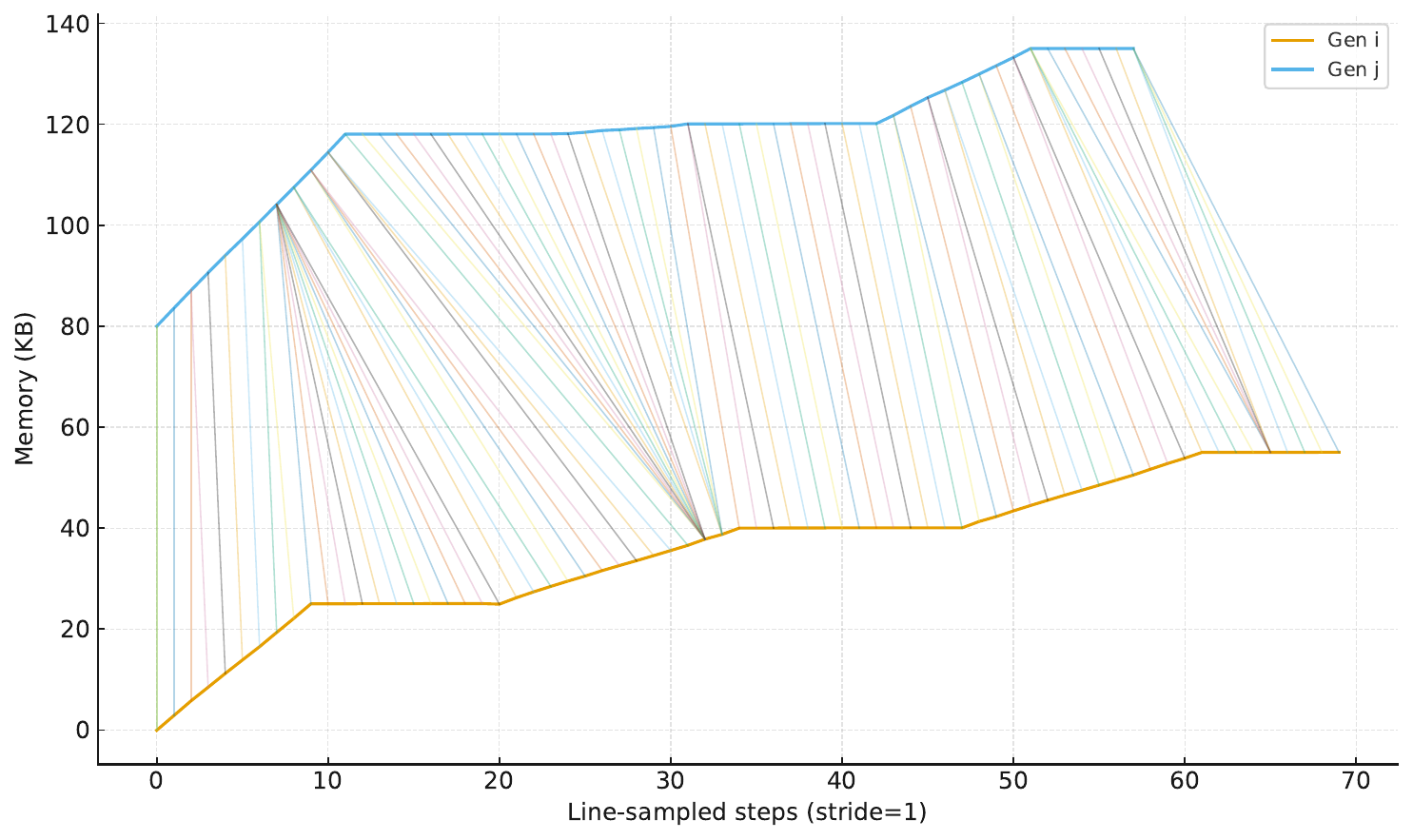}
    \vspace{-0.4em}
    \caption{Example: DTW sequence alignment}
    \label{fig:dtw}
  \end{figure}

\begin{figure*}[t]
\centering
\begingroup

\footnotesize
\setlength{\FrameRule}{0.45pt} 
\setlength{\FrameSep}{1.1pt}
\setlength{\OuterFrameSep}{0pt}

\def\SlideLeftW{0.43\textwidth}
\def\SlideRightW{0.555\textwidth} 

\setlength{\SlidingMaxColH}{6.25cm}

\def\SlideCodeSize{\scriptsize} 

\setlength{\abovecaptionskip}{2pt} 
\setlength{\belowcaptionskip}{0pt} 

\setlength{\parindent}{0pt}
\setlength{\parskip}{0pt}
\vspace{-0.25em}

\begin{minipage}[t][\SlidingMaxColH][t]{\SlideLeftW}

\begin{framed}
\textbf{Prompt}\par\vspace{1pt}
Given an array \texttt{nums} and window size \(k\ge1\), return a list where the \(i\)-th element is
\(\max(\texttt{nums}[i:i+k])\) for all valid windows.
\end{framed}

\vspace{0.8mm}

\begin{framed}
\textbf{Correct Solution A}\par\vspace{1pt}
\begin{lstlisting}[style=pyintro,
  basicstyle=\ttfamily\SlideCodeSize,
  aboveskip=0pt, belowskip=0pt,
  xleftmargin=0pt]
def sliding_max(nums: list[int], k: int) -> list[int]:
    # Builds (n - k + 1) overlapping slices of size k
    windows = [nums[i:i+k] for i in range(len(nums) - k + 1)]
    return [max(w) for w in windows]
\end{lstlisting}
\end{framed}

\vfill
\end{minipage}\hfill
\begin{minipage}[t][\SlidingMaxColH][t]{\SlideRightW}

\begin{framed}
\textbf{Correct Solution B}\par\vspace{1pt}
\begin{lstlisting}[style=pyintro,
  basicstyle=\ttfamily\SlideCodeSize,
  aboveskip=0pt, belowskip=0pt,
  xleftmargin=0pt]
from collections import deque

def sliding_max(nums: list[int], k: int) -> list[int]:
    dq = deque() # store indices, values decreasing in dq
    out = []
    for i, x in enumerate(nums):
        while dq and dq[0] <= i - k:      # drop indices out of window
            dq.popleft()
        while dq and nums[dq[-1]] <= x:   # maintain decreasing order
            dq.pop()
        dq.append(i)
        if i >= k - 1:
            out.append(nums[dq[0]])
    return out
\end{lstlisting}
\end{framed}

\vfill
\end{minipage}

\vspace{-6.5em} 
\caption{\centering Example of correct solutions with runtime divergence because of different space algorithmic strategies. Solution A on the left uses (\(O(nk)\) space while Solution B on the right uses \(O(k)\).}
\label{fig:slidingmax}

\endgroup
\end{figure*}

To isolate contestant code while suppressing allocator churn, we instrument with \texttt{tracemalloc} and convert line-sampled current-bytes into a \emph{Monotonic Peak Profile (MPP)}, the cumulative maximum of the baseline-corrected series (\autoref{fig:mpp_profiles_execution_memory}). This transform removes downward spikes from transient frees and GC cycles, yielding a non-decreasing envelope that highlights reproducible peak-growth events across runs and environments. We then compare unit-peak MPPs by \emph{shape}, using time-elastic Dynamic Time Warping (DTW) rather than raw magnitudes.

\begin{figure*}[t]
  \centering
  \includegraphics[page=1,width=0.7\textwidth]{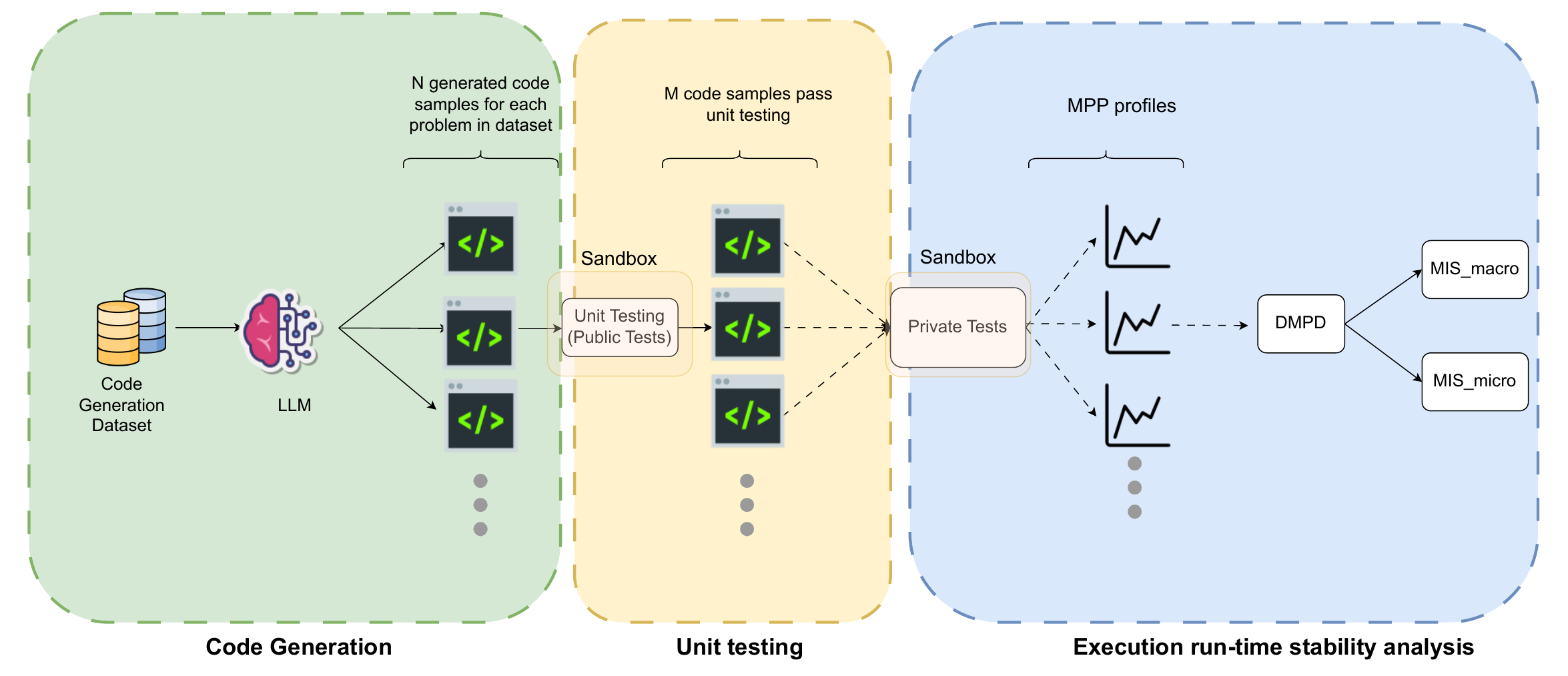}
  \caption{Pipeline for code generation and memory profiling}
  \label{fig:pipeline}
  \vspace{-1em}
\end{figure*}

\paragraph{Why runtime \emph{stability} matters.}
Temporal instability in execution-time memory profiles is not noise; it reflects underlying \emph{algorithmic structure} (Refer to Figure~\ref{fig:slidingmax} for an example case), choices of data structures, buffering, copy patterns, etc., even when solutions are functionally correct. By contrast, widely used downstream similarity metrics compare surface forms or static structure: CodeBLEU augments n\mbox{-}gram overlap with syntax and dataflow cues \cite{ren2020codebleu}, while AST-based measures such as Tree Structured Edit Distance (TSED/TED) quantify structural edits between program trees \cite{song2024revisiting}. These are informative for code similarity, but they do \emph{not} assess whether multiple correct generations from the \emph{same} model exhibit consistent behavior at runtime. Yet this is a fundamental aspect of real-world software engineering: consistency drives cost, capacity planning, tail failures, and churn-maintainability. In this work, we fill that gap by measuring runtime stability directly via shape-aware, scale-robust comparisons of memory trajectories (DMPD) and aggregating them to model-level instability (MIS). We also report how these stability proxies relate to established maintainability indicators: Cyclomatic Complexity \cite{mccabe1976complexity}, Maintainability Index \cite{oman1992metrics,coleman1994using}, and Cognitive Complexity \cite{campbell2018cognitive} to ground operational variability in standard SE practice.

\subsection{Industry Relevance and Scope}

Much like the well-documented security vulnerabilities in generated code, which have necessitated the standardization of rigorous security scanning prior to acceptance \cite{pearce2025asleep,perry2023users}, operational instability constitutes a parallel risk that hinders rapid adoption. In traditional container orchestration, deployment manifests mandate precise resource specifications (requests and limits) to govern bin-packing efficiency and horizontal autoscaling \cite{verma2015large}. Because these operational envelopes are tuned to historical baselines, an LLM-generated patch that introduces unmeasured memory variance can silently invalidate capacity estimates, forcing operators to determine if the new solution respects existing constraints or requires expensive re-tuning. If we can design methodologies that quantify the operational risk and cost of implementing LLM-generated patches in production systems, it becomes far easier to integrate AI agents into real delivery pipelines. Our work is one such step toward that goal, offering a stability-focused signal for comparing correct patches.

We use competitive programming benchmarks, which may differ materially from industrial codebases, as they would typically avoid deep dependency graphs, long-lived architectural constraints, configuration drift, concurrency, and ecosystem-level integration testing. These differences limit direct claims about production outcomes from our work. However, they also offer a controlled lens for isolating confounding factors when analyzing stability: competitive settings allow us to compare multiple correct generations under fixed inputs, making it easier to disentangle algorithmic choice, caching behavior, allocator/GC effects, and other low-level runtime phenomena from the noise of evolving microservices and environment-specific deployment quirks. Importantly, the underlying substrate that shapes allocation dynamics, such as language runtimes, memory allocators, and garbage collection, remains the same class of mechanism that production systems depend on. From a software engineering perspective, capacity planning and safe change management benefit from metrics that capture stability beyond correctness \cite{beyer2016site,beyer2018site,dean2013tail}. We therefore position our competitive-programming results as a methodological ``unit test'' for instability effects that are likely to reappear in noisier industrial settings, motivating follow-up validation on large-scale industry use cases.

\textbf{Our contributions are as follows.}
\vspace{-0.7em}
\begin{itemize}[leftmargin=*]
  \item \textbf{Instability among correct generations.} We show that LLM-generated programs exhibit \emph{runtime-memory} instability even when all solutions are functionally correct, and we quantify it with bounded metrics.
  \item \textbf{Temperature widens behavior (pass@1 vs.\ stability trade-off).} We show that raising sampling temperature consistently increases divergence as measured by \textsc{DMPD} and \textsc{MIS}, while often \emph{also} improving pass@1, revealing a controllable trade-off between functional success and runtime stability.
  \item \textbf{Robust metrics for fair comparison.} We provide the research community with metrics (\textsc{DMPD} and \textsc{MIS}) that are robust to confounding variables such as test-case magnitude and composition. By isolating the underlying algorithmic memory structure, our metrics enable fair comparisons of program stability across diverse input scales and test suites.
  \item \textbf{A link to software maintainability.} We connect our operational stability proxies to established software engineering indicators.
\end{itemize}

%% file: sections/research_questions.tex
\section{Research Questions}

\begin{enumerate}[label=\textbf{RQ\arabic*}]
  \item \textbf{Memory usage divergence.} To what extent do LLM-generated functionally correct programs generated for the same problem exhibit divergent runtime memory behaviors?
  \item \textbf{Impact of temperature on memory usage divergence.} How does the sampling temperature influence the consistency of runtime memory behavior across generated solutions?
  \item \textbf{Comparison with other metrics and baselines.} Do execution shape-aware metrics capture a novel dimension of code compared to other approaches and baselines like Normalized Peak Difference?
  \item \textbf{Robustness to scale and instrumentation.} To what degree are the instability measurements derived from our metrics influenced by variations in input scale and profiling configurations?
  \item \textbf{Relationship between operational stability and code quality.} What is the nature and strength of the association between runtime memory stability, as measured by \textsc{DMPD} and \textsc{NMV}, and established software engineering metrics of code quality?
\end{enumerate}

%% file: sections/approach.tex
\section{Approach}
Figure~\ref{fig:pipeline} shows the overview of our approach. We sample multiple correct solutions, run them to obtain memory-usage execution profiles (converted to monotonic peak profiles for robustness), and use Dynamic Time Warping to measure shape overlap. We define each component in detail in this section.

\subsection{Dynamic Time Warping}
\emph{Dynamic Time Warping} (DTW) is a classic technique that aligns sequences in time\cite{berndt1994using,keogh2005exact}. DTW treats two time series like flexible rubber bands laid over each other and asks: what is the \emph{least total effort} needed to line up their shapes if we are allowed to gently stretch or compress time? At each alignment step (Refer Figure~\ref{fig:dtw} for an alignment example), we pay a small \emph{mismatch cost} for the difference in values where the two traces are currently lined up; DTW then finds the path through these steps that \emph{minimizes} the accumulated cost. Our \emph{Dynamic Mean Pairwise Distance (DMPD)} summarizes this by taking the \emph{average} mismatch along DTW’s best alignment between \emph{unit-peak–normalized} MPPs, yielding a bounded score in \([0,1]\). The unit-peak step intentionally removes scale so DMPD is driven by \emph{shape}: where and how the profile rises, plateaus, and peaks. In practice, this separates cases with the same maximum but very different dynamics (early spike vs.\ late surge), yielding a robust, shape-based, and scale-robust way to measure instability.
DTW’s intuition “align shapes, not timestamps” is well established across domains: it originated in speech recognition to handle variable speaking rates \cite{sakoe2003dynamic}, underpins classic time-series pattern matching and classification \cite{berndt1994using,keogh2005exact}, aligns performances to musical scores and human motion traces \cite{muller2007information}, and is routinely used to compare noisy biomedical signals such as ECGs \cite{senin2008dynamic,giorgino2009computing}. We borrow the same idea here: treat two passing solutions as two \emph{shapes} of memory evolution, and measure how much “time-bending effort” it takes to make them agree.

\subsection{Monotonic Peak Profile (MPP)}
\label{sec:mpp}
Let $s_t$ be the traced bytes at discrete steps $t=1,\dots,T$. After baseline correction,
\[
s'_t = \max(0,\, s_t - s_1),\quad t=1,\dots,T,
\]
we define the \textbf{Monotonic Peak Profile (MPP)} $P=\{p_1,\dots,p_T\}$ as the cumulative maximum:
\begin{equation}
 p_t = \begin{cases}
   0 & \text{if } t=1, \\
   \max(p_{t-1},\, s'_t) & \text{if } t>1~.
 \end{cases}
 \label{eq:mpp}
\end{equation}
MPP is non-decreasing, emphasizes peak growth, and dampens transient alloc/free churn, improving alignment stability across runs and environments.

\subsection{Pairwise Comparison via DTW on Unit-Peak Profiles}

To compare the memory profiles of two programs, we first normalize their respective MPPs to isolate the shape of their memory usage from the absolute magnitude. Given two MPPs, $P_a$ (of length $n$) and $P_b$ (of length $m$), we scale each to have a unit peak:
\[
\widehat{P}_a = \frac{P_a}{\max P_a} \quad \text{and} \quad \widehat{P}_b = \frac{P_b}{\max P_b}.
\]
If the peak memory of a profile is zero, its normalized counterpart is defined as an all-zero sequence.

Next, we quantify the dissimilarity between these unit-peak profiles, $\widehat{P}_a$ and $\widehat{P}_b$, using Dynamic Time Warping (DTW). Let $D(i,j)$ be the cumulative $L^1$ cost for aligning the first $i$ points of $\widehat{P}_a$ with the first $j$ points of $\widehat{P}_b$. This cost is computed via the recurrence:
\begin{equation}
D(i,j) = |\widehat{p}_{a,i} - \widehat{p}_{b,j}| + \min\!\big\{D(i-1, j-1), D(i-1, j), D(i, j-1)\big\},
\label{eq:dtw_recurrence}
\end{equation}
with boundary conditions $D(0,0)=0$ and $D(i,0)=D(0,j)=\infty$ for $i,j>0$.

After computing the full cost matrix, we identify the optimal warping path, $\pi^\star$, by backtracking from the final cell $(n,m)$. Let $|\pi^\star|$ denote the length of this path, which represents the total number of aligned point-pairs between the two profiles.

Finally, we define the \emph{pairwise Divergence in Memory Profile Dynamics (DMPD)} as the average alignment cost per step along this optimal path:
\begin{equation}
\mathrm{DMPD}(P_a, P_b) = \frac{D(n,m)}{|\pi^\star|}.
\label{eq:dmpd_pair}
\end{equation}
Since the input profiles $\widehat{P}_a$ and $\widehat{P}_b$ are normalized to the range $[0,1]$, the resulting $\mathrm{DMPD}$ value is also bounded in $[0,1]$, where a value of $0$ indicates that the profiles have identical shapes.

\subsection{Instability Aggregation}
\paragraph{Problem-level score.}
For problem $p$, let $S_p$ be the set of successful solutions and $T_p$ the set of private tests.
Let $\binom{S_p}{2}$ denote all \emph{unordered} pairs of distinct solutions.
For each test $t\in T_p$, we compute $\mathrm{DMPD}$ for every pair $(i,j)\in\binom{S_p}{2}$ using the corresponding unit-peak MPPs.
The \emph{problem-level} instability is the mean over pairs and tests:
\begin{equation}
D_p \;=\; \frac{1}{\left|\binom{S_p}{2}\right|\cdot |T_p|}\;
\sum_{(i,j)\in \binom{S_p}{2}} \;\sum_{t\in T_p}\;
\mathrm{DMPD}\!\left(P_{p,i,t},\, P_{p,j,t}\right).
\label{eq:problem_score}
\end{equation}

\paragraph{Model Instability Score (MIS)}
Over a set of problems $\mathcal{P}$, we report the \emph{macro} MIS as the unweighted average of problem scores:
\begin{equation}
\mathrm{MIS}_{\text{macro}} \;=\; \frac{1}{|\mathcal{P}|}\sum_{p\in\mathcal{P}} D_p.
\label{eq:mis_macro}
\end{equation}
When problems vary widely in the number of successful solutions or tests, we additionally report a \emph{micro} average that weights by the number of evaluated pairs and tests:
\begin{equation}
\mathrm{MIS}_{\text{micro}} \;=\;
\frac{\sum_{p\in\mathcal{P}} \left|\binom{S_p}{2}\right||T_p|\, D_p}
{\sum_{p\in\mathcal{P}} \left|\binom{S_p}{2}\right||T_p|}.
\label{eq:mis_micro}
\end{equation}

\subsection{Normalized Maximum Velocity (NMV)}
From the MPP, $P=\{p_t\}_{t=1}^T$, define the per-step \emph{velocity}
\[
v_t \;=\; p_{t+1}-p_t \quad (t=1,\dots,T-1),
\]
and let $\mathrm{MaxVel}(P)=\max_t v_t$. 

For fairness across solutions on the \emph{same} private test, we normalize by that test's \emph{median peak} (computed across solutions that produced successful runs on the test). 
Let $P_{p,i,\ell}$ be the MPP for problem $p$, solution $i$, and test index $\ell$, and let 
$p^\star_{p,i,\ell}=\max P_{p,i,\ell}$ denote its peak. Define the test-level median peak
\[
\widetilde{p}_{p,\ell} \;=\; \operatorname{median}_{i \in S_p(\ell)} \big( p^\star_{p,i,\ell} \big),
\]
where $S_p(\ell)$ is the set of solutions that passed test $\ell$ for problem $p$. 
The \textbf{Normalized Maximum Velocity (NMV)} for $(p,i,\ell)$ is
\begin{equation}
\mathrm{NMV}_{p,i,\ell} \;=\; \frac{\mathrm{MaxVel}\!\left(P_{p,i,\ell}\right)}{\widetilde{p}_{p,\ell}}.
\label{eq:nmv_median}
\end{equation}
\noindent This per-test, across-solutions normalization yields comparable scales for thresholding within a test
(unlike peak-normalization by $p^\star_{p,i,\ell}$, \eqref{eq:nmv_median} is not necessarily $\le 1$; optionally, one may clip to $[0,1]$ if a hard bound is desired.).

\paragraph{Per-solution aggregation across private tests.}
We optionally restrict to \emph{eligible} runs with $p^\star_{p,i,\ell}\ge P_{\min}$ (e.g., $P_{\min}=100\,\mathrm{KiB}$). 
Let $T'_{p,i}\subseteq T_p$ be the eligible tests for solution $i$ on problem $p$. We report
\begin{align}
\overline{\mathrm{NMV}}_{p,i}
&= \frac{1}{|T'_{p,i}|}\sum_{\ell\in T'_{p,i}} \mathrm{NMV}_{p,i,\ell}
\end{align}
where $\tau>0$ is a configurable threshold. $\overline{\mathrm{NMV}}_{p,i}$ summarizes the average burstiness (relative to the test’s typical peak).

\iffalse
\subsection{Scaling and Single-Bin Sampling}
We stratify generated tests by \emph{magnitude} so each run contains a homogeneous scale class.

\paragraph*{Scale of a candidate.}
Let a generator return $r$ be coerced to arguments $A(r)=(a_1,\dots,a_m)$ (lists/tuples/maps recursively expanded). We define the numeric scale
\[
\sigma(r)=\max\{\,|x|:\ x\in A(r),\ x\in\mathbb{R}\,\},\qquad \max\varnothing:=0,
\]
and a fallback based on serialized input length $L(r)$:
\[
\widetilde{\sigma}(r)=
\begin{cases}
\sigma(r), & \sigma(r)>0,\\
L(r), & \sigma(r)=0~.
\end{cases}
\]

\paragraph*{Log bins (base-10).}
With base $\beta=10$ and starts $S=\{1,10,100,1000\}$, define half-open bins
\[
B(s)=[\,s,\ \beta s\,)= [\,s,\ 10s\,),\qquad s\in S.
\]
Edge cases follow the interval convention: $s\le \widetilde{\sigma}(r)<10s$.

\paragraph*{Single-bin selection (one run).}
Fix a \emph{bin start} $s^\star\in S$ and a \emph{count} $c\in\mathbb{N}$. From the candidate stream $\mathcal{R}$:
\begin{align}
\mathcal{C}(s^\star) &= \big\{\, r\in\mathcal{R}\ :\ \widetilde{\sigma}(r)\in B(s^\star)\ \wedge\ I(r)\ \text{unseen}\,\big\},\\
\mathrm{Out}(s^\star,c) &= \text{First}_c\!\big(\operatorname{Order}_\seed(\mathcal{C}(s^\star))\big),
\end{align}
where $I(r)$ is the serialized input string used for deduplication and $\operatorname{Order}_\seed(\cdot)$ is a fixed seeded ordering (deterministic across runs with the same seed).
\fi

Beyond the formal definitions, the Algorithm below summarizes the approach in pseudocode.

\par\smallskip
\noindent
\begingroup

\footnotesize
\setlength{\FrameSep}{0.8pt}
\setlength{\parskip}{0pt}
\setlength{\topsep}{0pt}
\setlength{\partopsep}{0pt}
\setlength{\parsep}{0pt}
\setlength{\itemsep}{0pt}
\setlength{\algorithmicindent}{1.1em}

\begin{Algobox}{Algorithm A: Code Gen $\rightarrow$ Unit Tests $\rightarrow$ DMPD Tables}
\begin{algorithmic}
\STATE \textbf{Dataset:} problems (\texttt{problem\_id}, desc, public/private tests);
       \textbf{Settings:} models $\mathcal{M}$, temps $\mathcal{T}$, completions $N$, private test cap $r$
\STATE \textbf{A. Code generation dataset (JSONL)}
\FOR{$m \in \mathcal{M}$, $\tau \in \mathcal{T}$}
  \FOR{problem $P$}
    \FOR{$i = 1$ \TO $N$}
      \STATE prompt LLM with $P$ at $(m,\tau)$; extract Python block
      \STATE write JSON (\texttt{problem\_id}, \texttt{solution\_id}, $i$, code, tests, raw)
    \ENDFOR
  \ENDFOR
\ENDFOR
\STATE \textbf{B. Public unit tests (filter)} run test sandbox; label each record \texttt{success}/\texttt{fail}
\STATE \textbf{C. Private tests \& profiling (per problem)}
\FOR{problem $p$}
  \STATE $S_p \leftarrow$ solutions labeled \texttt{success}
  \FOR{solution $i \in S_p$}
    \FOR{private test $\ell \le r$}
      \STATE compile with filename \texttt{\textless contestant\textgreater}
      \STATE run with timeout; compare output; \textbf{skip} on fail/timeout/mismatch
      \STATE \emph{tracemalloc}: sample \emph{current bytes} restricted to \texttt{\textless contestant\textgreater}
      \STATE baseline subtract first sample; floor at $0$; optional quantization $q$
      \STATE transform samples $\rightarrow$ \textbf{MPP} (cumulative max); record peak/max velocity
    \ENDFOR
  \ENDFOR
  \STATE \textbf{Pairwise distances (per test)}
  \FOR{private test $\ell$}
    \FOR{each unordered pair $(i,j)$ with valid MPPs}
      \STATE unit-peak normalize both MPPs
      \STATE compute \textbf{DMPD} via DTW with $L^1$ local cost (avg cost along optimal path)
      \STATE append \texttt{(problem $p$, test $\ell$, pair $(i,j)$, DMPD)} to table
    \ENDFOR
  \ENDFOR
\ENDFOR
\STATE \textbf{Output:} per-problem, per-test pairwise DMPD tables (no aggregation)
\end{algorithmic}
\end{Algobox}

\vspace{-5pt}

\begin{Algobox}{Algorithm B: Aggregation to MIS\_macro and MIS\_micro}
\begin{algorithmic}
\STATE \textbf{Inputs:} DMPD tables from Algorithm~A
\STATE \textbf{A. Problem-level scores}
\FOR{problem $p$}
  \STATE gather all DMPD values across tests and unordered pairs
  \STATE $\#\text{pairs}_p \leftarrow$ evaluated pairs; $\#\text{tests}_p \leftarrow$ evaluated tests
  \STATE \textbf{$D_p$} $\leftarrow$ mean of DMPD values for $p$
  \STATE \textbf{$w_p$} $\leftarrow \#\text{pairs}_p \times \#\text{tests}_p$
\ENDFOR
\STATE \textbf{B. Cross-problem aggregation}
\STATE \textbf{MIS\_macro} $\leftarrow$ mean of $D_p$ for problems with $w_p > 0$ (equiv.: mean of all DMPD entries pooled across problems)
\STATE \textbf{MIS\_micro} $\leftarrow$ weighted mean of $D_p$ using weights $w_p$
\STATE \textbf{Output:} table $(p, D_p, w_p)$ and global MIS\_macro, MIS\_micro
\end{algorithmic}
\end{Algobox}

\endgroup
\par\smallskip

%% file: sections/experiments.tex
\section{Experiments}

\begin{table}[h!]
  \centering
  \setlength{\aboverulesep}{0pt}
  \setlength{\belowrulesep}{0pt}
  \renewcommand{\arraystretch}{0.9}
  \setlength{\tabcolsep}{2pt}
  \scriptsize
  \caption{Large language models selected for our stability evaluation, grouped by source and specialization. Abbreviations are used in plots and figures.}
  \vspace{-1em}
  \label{tab:model_selection}
  \begin{tabular}{@{} l l l l l @{}}
    \toprule
    \textbf{Model} & \textbf{Abbreviation} & \textbf{Params} & \textbf{Context} & \textbf{Developer} \\
    \midrule

    \multicolumn{5}{@{}>{\columncolor{gray!20}}l}{\textbf{Commercial Models}} \\
    \multicolumn{5}{@{}l}{\textbf{\textit{Language/Code Models}}} \\
    GPT-3.5-turbo-instruct     & GPT-3.5       & N/A   & 16k   & OpenAI \\
    GPT-4o                     & GPT-4o        & N/A   & 128k  & OpenAI \\
    Claude-3.7-Sonnet          & Claude-3.7-S  & N/A   & 200k  & Anthropic \\

    \multicolumn{5}{@{}l}{\textbf{\textit{Reasoning Model}}} \\
    GPT-o4-mini                & GPT-o4-m      & N/A   & 128k  & OpenAI \\

    \midrule
\multicolumn{5}{@{}>{\columncolor{gray!20}}l}{\textbf{Open-Source Models}} \\
    \multicolumn{5}{@{}l}{\textbf{\textit{Code Models}}} \\
    Qwen2.5-Coder-7B           & Qwen-7B-C     & 7B    & 64k   & Alibaba Cloud \\
    CodeLlama-7B-Instruct      & CodeLlama-7B-It & 7B  & 16k   & Meta \\
    Codestral-22B              & Codestral-22B & 22B   & 32k   & Mistral AI \\

    \multicolumn{5}{@{}l}{\textbf{\textit{Language/Code Models}}} \\
    Llama-3.1-8B               & Llama3.1-8B   & 8B    & 128k  & Meta \\
    Mistral-7B-v0.3            & Mistral-7B    & 7B    & 32k   & Mistral AI \\

    \multicolumn{5}{@{}l}{\textbf{\textit{Reasoning Models}}} \\
    DeepSeek-R1-Distill-Qwen-32B & DS-Qwen-32B & 32B   & 128k  & DeepSeek AI \\
    DeepSeek-R1-Distill-Llama-70B & DS-Llama-70B & 70B & 128k  & DeepSeek AI \\

    \bottomrule
  \end{tabular}
  \vspace{-2em}
\end{table}
\noindent\textbf{Empirical Protocol.}
Our empirical protocol is designed to systematically quantify the \emph{runtime memory} stability of code artifacts generated by LLMs. The fundamental task under investigation is code generation: for each experimental run, the input is a prompt containing a complete problem description drawn from our benchmarks (formal statement, I/O specification, and constraints), and the expected output is a compilable Python function intended to solve the given problem. For each problem–model pair, we generate a corpus of candidate solutions (n = 5 in our experiments) to analyze their collective stability. A critical prerequisite for this analysis is the establishment of \emph{functional correctness}. We deem a generated artifact correct if it passes all public unit tests provided by the source benchmark. Our stability study is therefore conditioned on correctness. Concretely, we profile memory-allocation dynamics for the cohort of correct solutions, transform raw traces into \emph{Monotonic Peak Profiles} (MPP; Eq.~\ref{eq:mpp}), and compare unit-peak profiles with Dynamic Time Warping to obtain \emph{DMPD} (Eq.~\ref{eq:dmpd_pair}), which we aggregate to per-problem scores and cross-problem \(\mathrm{MIS}_{\text{macro}}\) and \(\mathrm{MIS}_{\text{micro}}\) (Eqs.~\ref{eq:problem_score}–\ref{eq:mis_micro}). 

We also compute static maintainability metrics: Cyclomatic Complexity \cite{mccabe1976complexity}, Cognitive Complexity \cite{campbell2018cognitive}, and Maintainability Index \cite{oman1992metrics,coleman1994using} and join them with the per\mbox{-}solution \(\overline{\mathrm{DMPD}}\) (mean over private tests; likewise \(\overline{\mathrm{NMV}}\)). We report Pearson/Spearman correlations across all solutions (Table~\ref{tab:se_from_stability_one_table} and visualize distributional shifts with double violin plots (Figure~\ref{fig:violen_plot_with_SE}) by stratifying \(\overline{\mathrm{DMPD}}\)/\(\overline{\mathrm{NMV}}\) into tertiles (T1 low \(\rightarrow\) T3 high), alongside Cliff’s \(\delta\) for T1 vs.\ T3.

\subsection{Datasets and Models}
To comprehensively evaluate our approach, we utilize two distinct problem suites: \textsc{CodeContests} and \textsc{BigOBench}. \textsc{CodeContests}, with its 165 problems and graded difficulty, serves as a challenging benchmark for functional correctness. In contrast, \textsc{BigOBench} is specifically designed to probe the link between algorithmic complexity and runtime behavior; we use its 318-problem space-complexity test set. A key feature of both datasets is their inclusion of public and private unit tests. This structure allows us to first identify functionally correct solutions using the public tests and then analyze their dynamic behavior on the private ones.
On these benchmarks, we evaluate a diverse set of eleven Large Language Models (LLMs) to ensure the generality of our findings (see Table~\ref{tab:model_selection}). These models, spanning both commercial and open-source systems, are grouped into three functional categories: (1) \textbf{code-generation} models fine-tuned for synthesis, (2) general-purpose \textbf{language/code} models, and (3) \textbf{reasoning} models optimized for multi-step problem solving. This curated selection enables a robust analysis across various model scales, architectures, and specializations.

\paragraph{Mercury dataset.} %(for scaling analyses).
To study scale effects on \(\mathrm{MIS}_{\text{macro}}\), \(\mathrm{MIS}_{\text{micro}}\), and \(\mathrm{DMPD}\), we use the \textsc{Mercury} dataset~\cite{du2024mercury}, whose test set contains 256 problems. \textsc{Mercury} provides per-problem test-case generators, which we exploit to synthesize private tests in four strict “size” buckets \(N\in\{1, 10, 100, 1000\}\). Here, \(N\) denotes the characteristic input magnitude for the primary type (e.g., list length, string length, or number of rows for list-of-lists/matrices, depending on the problem). For each problem and bucket, we generate 10 private tests. Implementation details follow a strict-size sampler: we infer an input-serialization profile from the public tests, generate candidates via the problem’s \texttt{generator\_code}, clamp serialized inputs to the target size window, deduplicate by serialized form, and deterministically select the first \(K\) examples. Expected outputs are obtained by executing an accepted in-file reference solution.

\subsection{Metric Calculation and Execution Protocol}
Our experimental protocol begins with generating \(N{=}5\) solutions for each model–problem pair across three temperatures: \(\{0.0,\,0.7,\,0.95\}\). This selection allows us to span a \emph{near-deterministic baseline} (\(0.0\)), a \emph{moderate-diversity} setting widely used for code assistants (\(\approx 0.7\)~\cite{schick2023toolformer}), and a \emph{high-diversity} regime (\(0.95\)) that accentuates solution variation useful for probing the pass@1 vs.\ stability trade-off. To examine temperature effects more finely, we additionally perform a systematic temperature scan for \texttt{gpt-3.5-turbo-instruct} from \(0.0\) to \(2.0\). Each generated solution is then executed against up to \(r{=}10\) private tests, using a single-bin input-magnitude policy with a fixed seed to ensure each run occupies a comparable scale regime. During execution, we instrument application-level memory via \texttt{tracemalloc} (\emph{current bytes}) and scope attribution to the contestant module. Execution is guarded by per-run timeouts; runs that fail (timeout, OOM, mismatch, or instrumentation error) are excluded from aggregation but retained in logs. The collected memory traces are then transformed into MPPs (Eq.~\ref{eq:mpp}) after baseline-subtraction and quantization. From these MPPs, we compute pairwise distances as \emph{DMPD} (Eq.~\ref{eq:dmpd_pair}) using DTW with an \(L^{1}\) local cost. Finally, these scores are aggregated to form the per-problem instability metric \(D_p\) (Eq.~\ref{eq:problem_score}) and the cross-problem aggregates \(\mathrm{MIS}_{\text{macro}}\) and \(\mathrm{MIS}_{\text{micro}}\) (Eqs.~\ref{eq:mis_macro}–\ref{eq:mis_micro}).

\noindent\emph{Practical note on measurement noise.}
Even with application-level scoping, memory traces reflect interpreter and allocator effects (reference counting, GC cycles), library behavior, scheduling, and incidental system activity. Our MPP transform and unit-peak, time-elastic alignment substantially reduce, but cannot eliminate this variability. Accordingly, our defaults prioritize stability: \emph{line sampling} with stride \(s{=}1\), byte quantization \(q{=}64\)~B, strict filename scoping, and unconstrained DTW with \(L^{1}\) costs; we report distributional summaries (medians/IQRs) and sensitivity ablations (including time sampling with \(\Delta t\in\{0.1,1,2\}\,\mathrm{ms}\) and coarser \(q\)) to separate model-induced divergence from residual profiling noise.

\subsection{Profiling Hyperparameters}
\label{sec:hyperparams}
We convert raw memory traces into stable profiles using two hyperparameters that control quantization and sampling. To reduce allocator noise, we \textbf{quantize} memory by rounding each sample to the nearest multiple of \(q\) bytes (default \(q{=}64\)\,B); larger \(q\) yields smoother profiles but lower resolution. To align memory with program structure rather than wall-clock effects, we sample memory every \(s\)-th executed line in the contestant's module (\textbf{line stride}, default \(s{=}1\)), which reduces sensitivity to machine speed. We compared this to time-based sampling and found line-based sampling more stable. Unless stated otherwise, we report results with \(s{=}1\) and \(q{=}64\)\,B.

\iffalse
\subsection{Strict-size private test sampling (single-bin)}
\label{sec:strict_size_sampler}
We generate private tests inside a \emph{single} serialized-input size window \([\,\texttt{size\_min},\,\texttt{size\_max})\) to remove scale effects. The pipeline is deterministic:
\begin{enumerate}[leftmargin=1.1em,itemsep=0.15em]
  \item Infer an I/O profile from public tests (arity/shapes, optional \texttt{convert\_offline}).
  \item Generate candidates, serialize inputs, keep those with length \(L\) s.t.\ \(\texttt{size\_min}\le L<\texttt{size\_max}\); deduplicate by the serialized string.
  \item With a fixed seed, take the first \(\texttt{num\_private}\) candidates.
  \item Solve outputs using an accepted in-file reference solution (fallback: guarded model completion).
\end{enumerate}
This strict-size, single-bin policy controls a major confound (test-size drift), so differences in \textsc{DMPD}/\textsc{MIS} reflect \emph{algorithmic memory shape}.
\fi

%% file: sections/results.tex
\section{Results} 

\subsection{Memory Usage Divergence}
\label{sec:results:operability}

\begin{figure}[htbp]
  \centering
  \footnotesize
  \includegraphics[page=1,width=\linewidth]{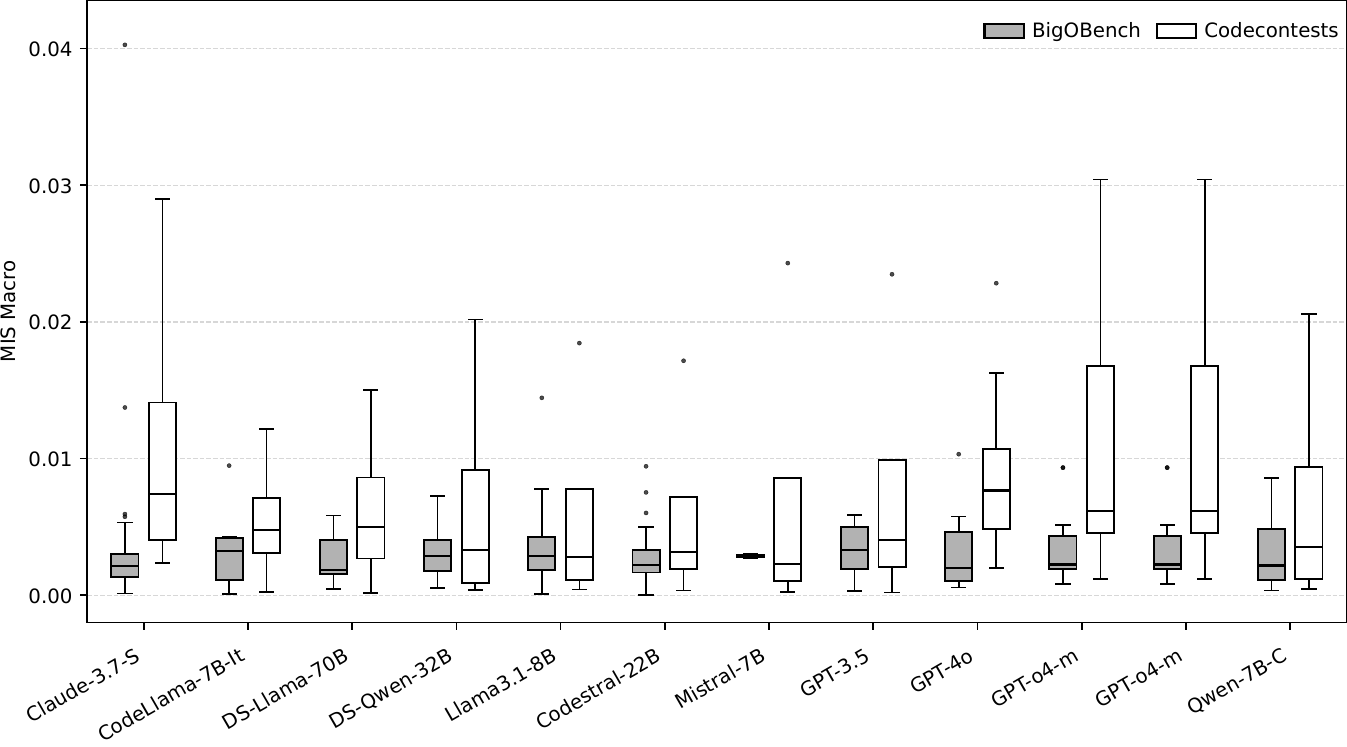}
  \caption{MIS macro for BigObench and CodeContests}
  \label{fig:mis_macro}
  \vspace{-1em}
\end{figure}

\begin{figure}[htbp]
  \centering
  \includegraphics[page=1,width=\linewidth]{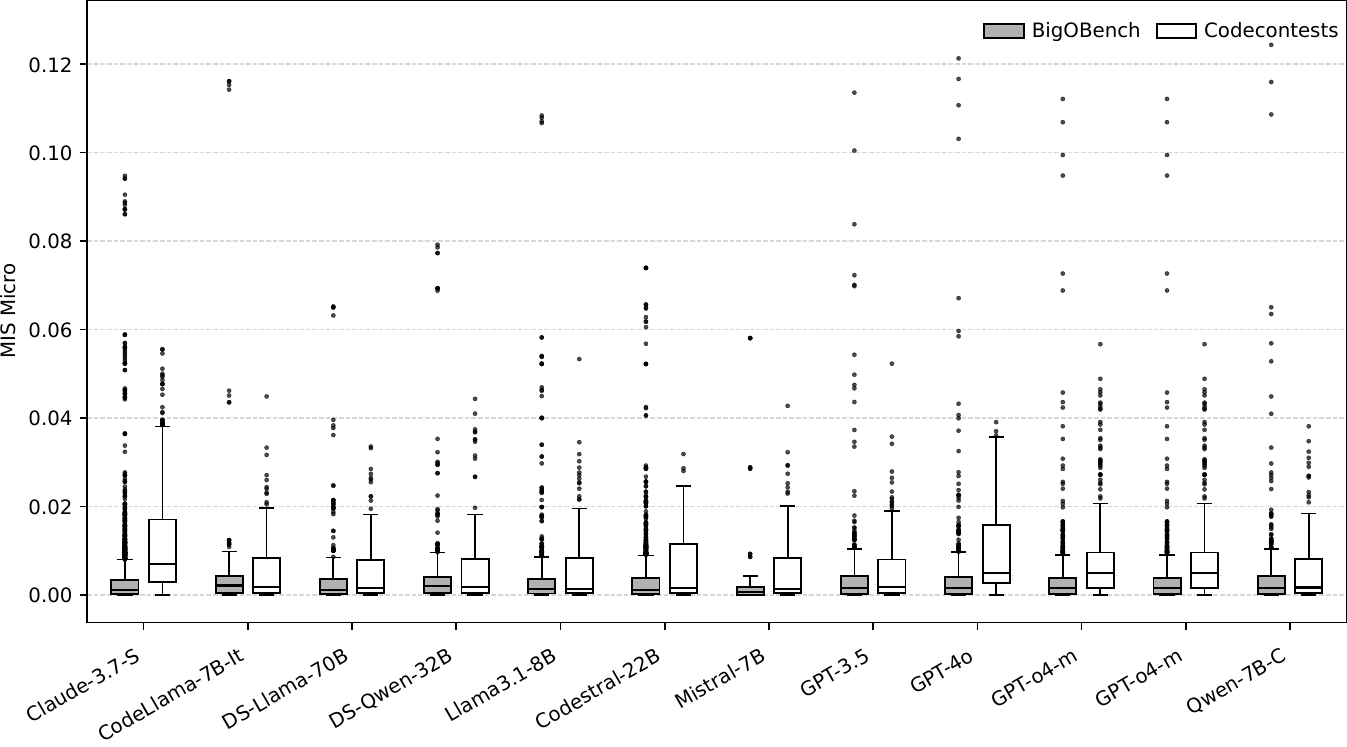}
  \caption{MIS micro for BigObench and CodeContests}
  \label{fig:mis_micro}
  \vspace{-2em}
\end{figure}

Restricting analysis to \emph{passing} solutions does not collapse behavior to a single runtime mode: for the same problem, we observe cohorts whose MPP envelopes either cluster tightly or diverge markedly (Figure~\ref{fig:mpp_profiles_execution_memory}). This dispersion is captured by non-zero pairwise \textsc{DMPD} values and, when aggregated, by \textsc{MIS}, which summarizes how much the shapes of memory growth differ among correct generations (\Cref{fig:mis_macro,fig:mis_micro}) for the complete dataset.

Crucially, \emph{functional} success and \emph{operational} stability decouple. Models with similar pass@1 can exhibit very different instability magnitudes. For example, on \textsc{BigOBench}, \texttt{GPT-4o} increases pass@1 from \(0.68\) (T{=}0) to \(0.80\) (T{=}0.95), while its \(\mathrm{MIS}_{\text{macro}}\) simultaneously rises from \(0.0042\) to \(0.0072\) (Table~\ref{tab:main_results_new}), indicating more varied memory-evolution shapes at higher temperature. By contrast, \texttt{Claude-3.7-Sonnet} attains higher pass@1 with a lower \textsc{MIS}, showing that two models with comparable correctness can differ materially in runtime stability.

\paragraph{Implication.}
In production, two correct programs can carry different memory costs. Instability among correct generations therefore impacts capacity planning; correctness thus is a necessary gate, not a sufficient guarantee of \emph{operability}. Stability-aware selection (e.g., reranking by \textsc{MIS}/\textsc{DMPD}) can reduce unseen real-world software engineering risks without sacrificing pass@1 (effect size on maintainability is discussed in Section\ref{sec: to_se}.

\begin{paperbox}[Answer to RQ1]{analysisbg}
\textbf{Among functionally correct solutions, we observe substantial divergence in memory trajectories}. Pairwise \textsc{DMPD} is non-zero and aggregate \textsc{MIS} varies by model and temperature (\Cref{fig:mpp_profiles_execution_memory,fig:mis_macro,fig:mis_micro}, Table~\ref{tab:main_results_new}). Correctness does not imply runtime stability.
\end{paperbox}

\subsection{Impact of Temperature on Memory Usage Divergence}
\label{sec:results:temperature}

\begin{figure}[htbp]
  \centering
  \includegraphics[page=1,width=\linewidth]{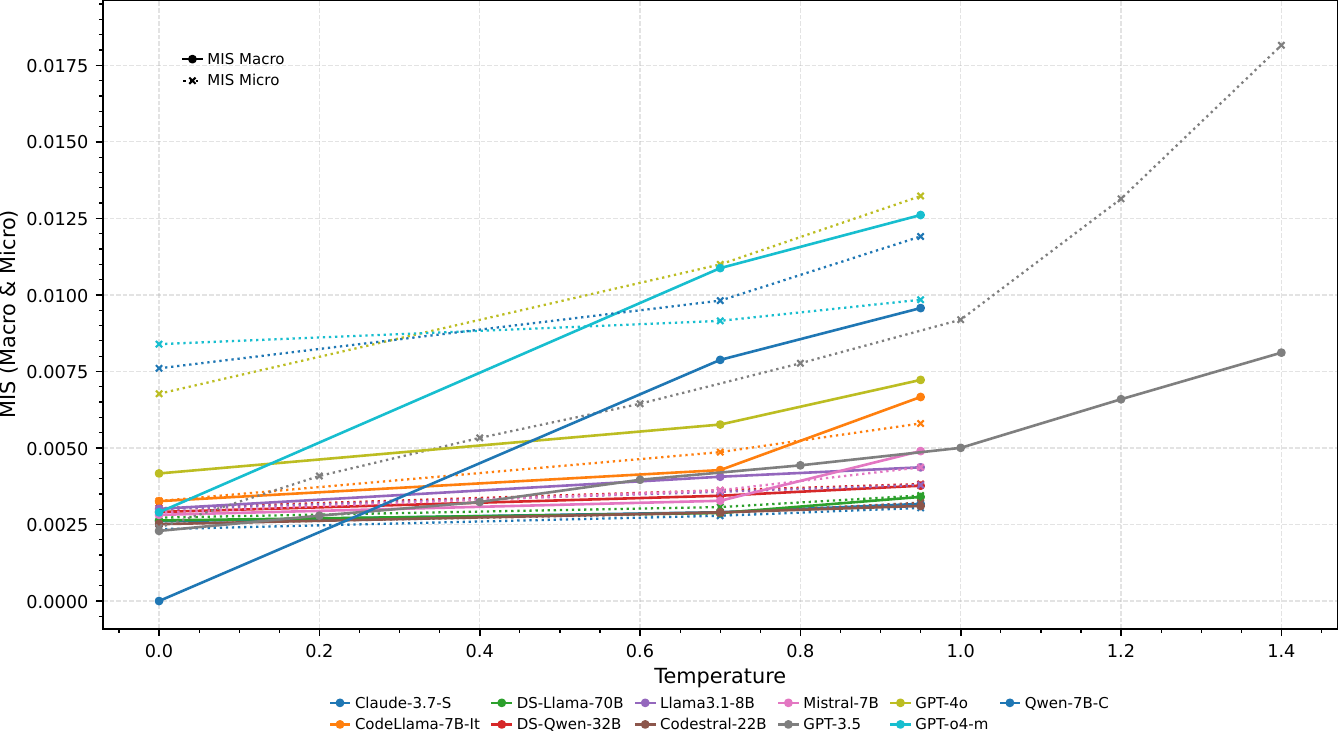}
  \caption{MIS macro and micro relation with temperature on BigOBench}
  \label{fig:Mis_with_temp}
\end{figure}

\newcommand{\tablescale}{1} 
Sampling temperature \(T\)—the standard knob for exploring an LLM’s internal solution space—is ubiquitous in practice; we therefore study how \(T\) shapes runtime \emph{stability} among correct generations. We consistently observe that an increase in temperature widens runtime behavior among correct generations. Across models and both suites, \(\mathrm{MIS}_{\text{macro}}\) and \(\mathrm{MIS}_{\text{micro}}\) generally increase with temperature (\Cref{fig:Mis_with_temp}). For instance, on \textsc{BigOBench}, \texttt{GPT-o4-mini} shows \(\mathrm{MIS}_{\text{macro}}\) growing from \(0.0029\) (T\(=0\)) to \(0.0126\) (T\(=0.95\)) while pass@1 also rises (0.69\(\rightarrow\)0.80), indicating temperature perturbs execution \emph{paths} (data structures, control flow), not just surface syntax. For \texttt{gpt-3.5-turbo-instruct}, we scan \(T\in[0,2]\); beyond \(T\approx1.4\) it does not yield \(\ge 2\) correct generations for any problem, precluding \textsc{DMPD}/\textsc{MIS} computation (which require at least two passing solutions).

\paragraph{Implication.}
Higher temperature explores a broader solution space: pass@1 may improve, but the same diversity amplifies runtime-memory instability (higher \textsc{DMPD}). 

\begin{paperbox}[Answer to RQ2]{analysisbg}
\textbf{Higher temperature increases instability.} Both \(\mathrm{MIS}_{\text{macro}}\) and \(\mathrm{MIS}_{\text{micro}}\) rise with sampling temperature across most model–dataset pairs (\Cref{fig:Mis_with_temp}, Table~\ref{tab:main_results_new}), reflecting broader solution-manifold exploration that diversifies memory dynamics.
\end{paperbox}

\subsection{Comparison With Other Metrics and Baselines}
\label{sec:results: orthogonality}
\begin{figure}[htbp]
  \centering
  \includegraphics[page=1,width=\linewidth]{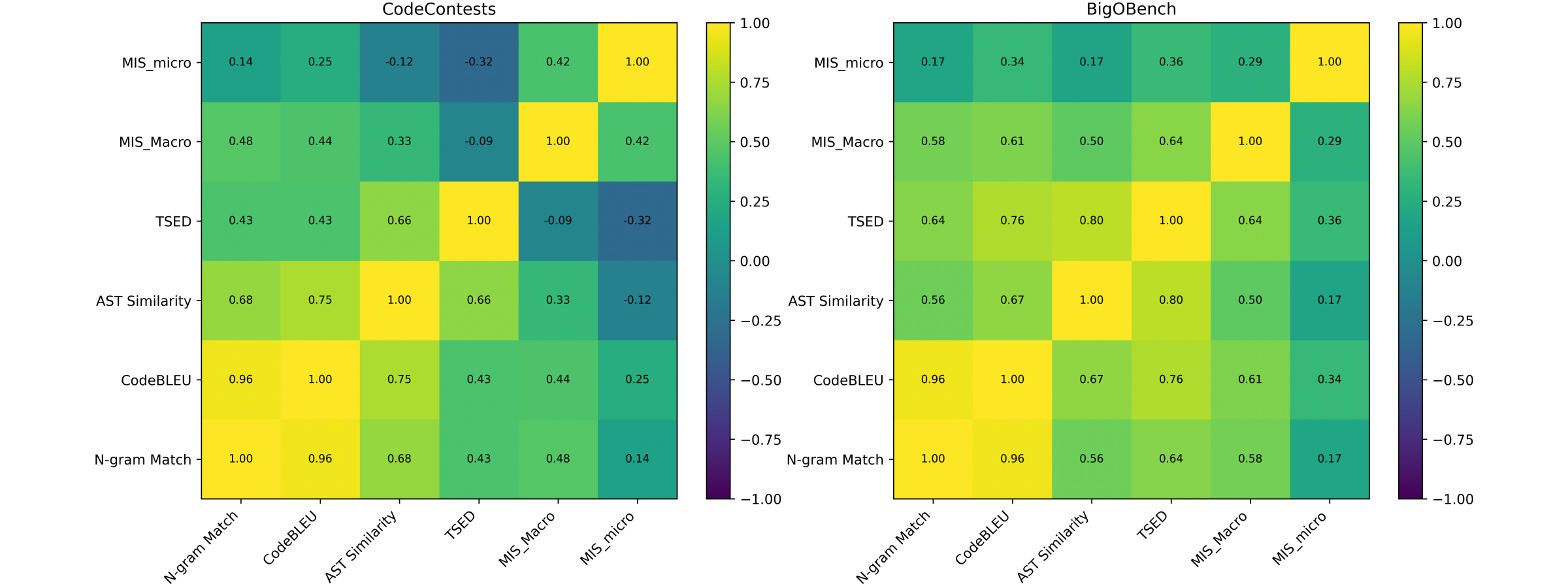}
  \caption{Pearson correlation heatmap for GPT-o4-mini on traditional metrics vs MIS on BigOBench}
  \label{fig:pearson_corr}
  \vspace{-1em}
\end{figure}

\textsc{MIS} is largely orthogonal to surface/syntactic similarity (n\mbox{-}gram/Code\\BLEU~\cite{ren2020codebleu}, and structure similarity (AST similarity,  TSED~\cite{song2024revisiting}). See correlation heatmap (Figure~\ref{fig:pearson_corr}), indicating that it captures a different, execution-time dimension of code behavior.

\begin{figure}[t]
  \centering
  \includegraphics[page=1,width=0.9\linewidth]{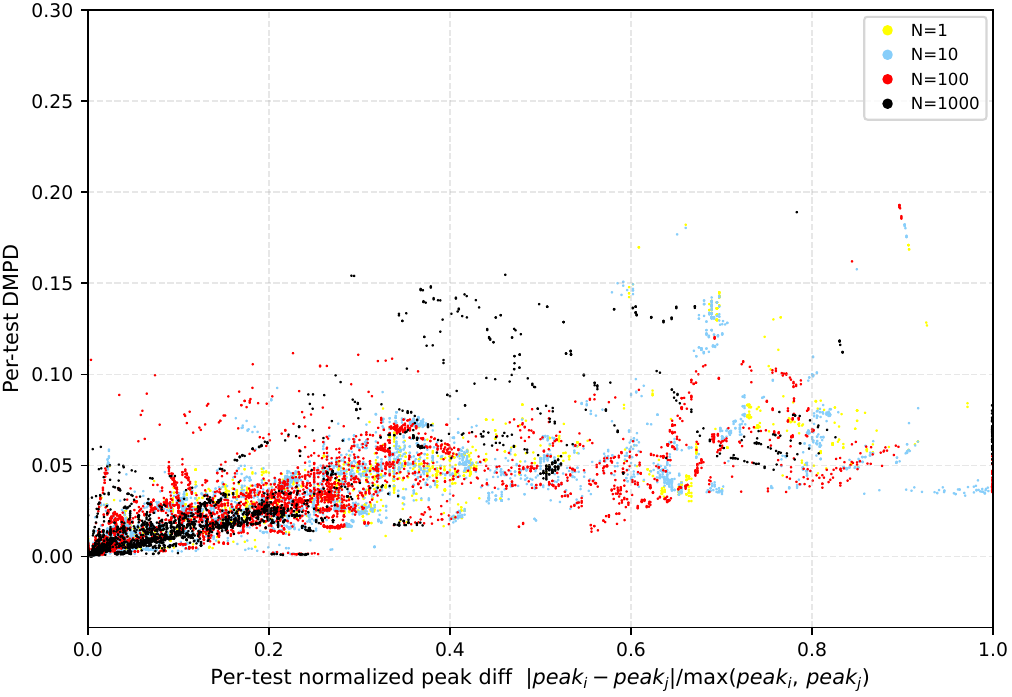}
  \caption{Openai-o4-mini DMPD vs Normalized peak difference with scale (on Mercury dataset)}
  \label{fig:dmpd_vs_normalized_peaks}
\end{figure}

Peak-only proxies such as normalized peak difference (NPD) vary widely with the test case and input scale. In \Cref{fig:dmpd_vs_normalized_peaks}, NPD shows large, test-dependent spread, whereas \textsc{DMPD} clusters tightly, reflecting its \emph{scale invariance}. Although we observe a strong but imperfect correlation (\(R^2\!\approx\!0.75\)) between per\mbox{-}problem mean NPD and \textsc{DMPD} i.e., solutions unstable in shape often also fluctuate in amplitude, the amplitude proxy is far more sensitive to the particular private tests, while \textsc{DMPD} is not. Consistently, across explicit scale factors \(N\in\{1,10,100,1000\}\), \Cref{fig:mis_with_scale} shows \textsc{MIS} remains fairly stable with scale, mirroring \textsc{DMPD}'s behavior (with additional damping from aggregation).

\paragraph{Implication.}
Shape\mbox{-}based, time\mbox{-}elastic metrics (\textsc{DMPD}/\textsc{MIS}) are more \emph{robust} indicators of runtime instability across test suites and input scales than peak\mbox{-}based proxies.

\begin{paperbox}[Answer to RQ3]{analysisbg}
\textbf{Orthogonal to traditional metrics and more robust than baseline.} \textsc{DMPD}/\textsc{MIS} capture orthogonal dimension missed by traditional metrics (\Cref{fig:pearson_corr}), and remain stable under input scaling (\Cref{fig:dmpd_vs_normalized_peaks,fig:mis_with_scale}), whereas peak-memory based baseline varies widely with specific tests. 
\end{paperbox}

\subsection{Robustness to Scale and Instrumentation}
\label{sec:results:measurement_bias}

\begin{figure}[t]
  \centering
  \includegraphics[page=1,width=0.9\linewidth]{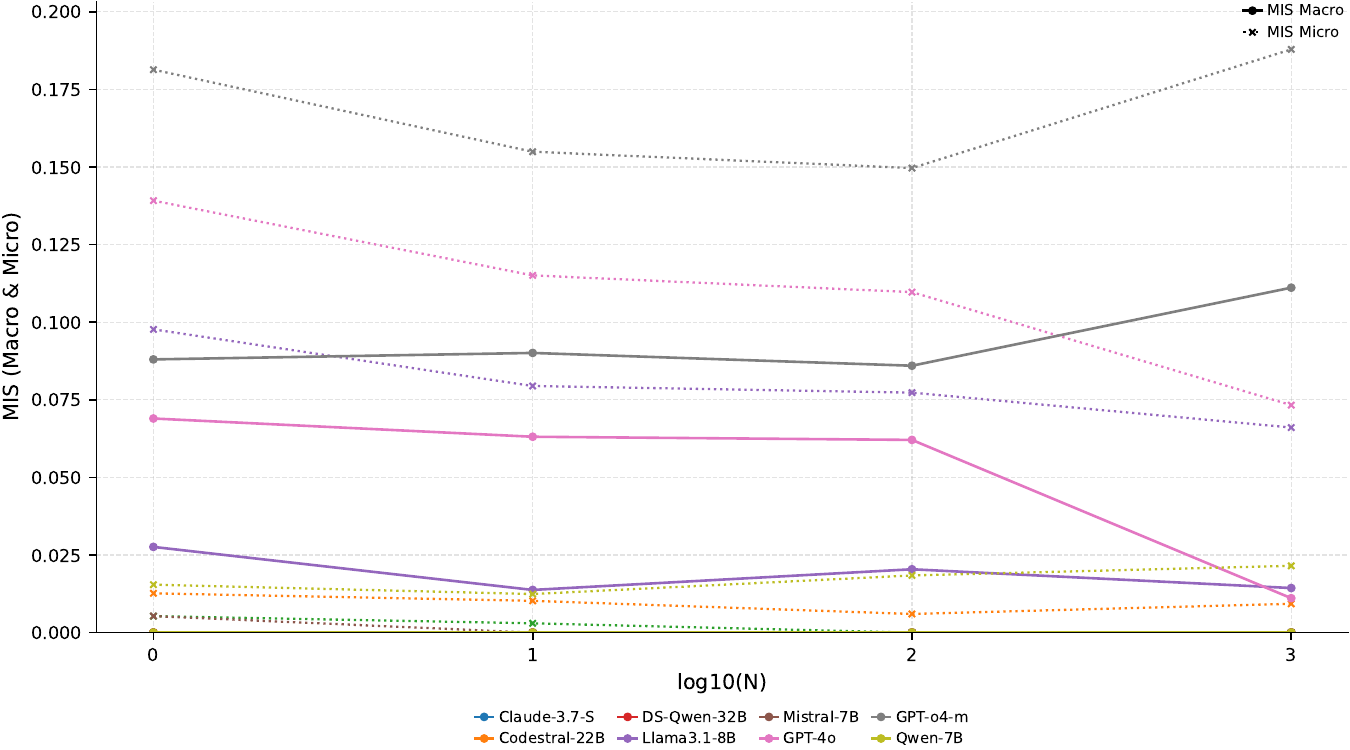}
  \caption{MIS micro and macro with scale (on Mercury dataset)}
  \label{fig:mis_with_scale}
  \vspace{-1em}
\end{figure}

Ablations show that \textsc{MIS} is fairly stable with scaled input sizes of test suites (Figure~\ref{fig:mis_with_scale}), but profiling choices can inflate or dampen perceived instability (\Cref{tab:hyperparam_sensitivity}). Relative to line sampling with stride \(s{=}1\) and 64\,B quantization (baseline), increasing stride to \(s{=}5\)–\(10\) modestly raises MIS (+5–9\%) and inflates dispersion. Coarsening byte quantization to 128–256\,B causes negligible (\(<\!1\%\)) effect. In contrast, fine-grained \emph{time} sampling at 0.1\,ms overstates instability (+46\% macro / +55\% micro), consistent with allocator/GC micro-effects~\cite{gcdocs}; 1–2\,ms reduces but does not eliminate the bias.

\paragraph{Implication.}
Instability estimates should reflect \emph{model/code} diversity, not measurement noise. Our defaults (line sampling, \(s{=}1\), modest quantization) provide conservative, reproducible estimates; time sampling is appropriate only when wall-clock constraints dominate and intervals are sufficiently coarse.

\begin{paperbox}[Answer to RQ4]{analysisbg}
\textbf{Conclusions are robust under reasonable settings, but measurement matters.} Prefer line sampling and modest quantization; avoid very fine time grids that amplify allocator noise (\Cref{tab:hyperparam_sensitivity}).
\end{paperbox}

\begin{table*}[t]
\centering
\caption{Comparative Analysis of LLMs: pass@1, MIS\_macro, and MIS\_micro across BigOBench and CodeContests}
\label{tab:main_results_new}
\resizebox{\tablescale\textwidth}{!}{%
\begin{tabular}{@{}lccccccccccccccccccc@{}}
\toprule
& \multicolumn{6}{c}{\textbf{pass@1}}
& \multicolumn{6}{c}{\textbf{MIS\_macro}}
& \multicolumn{6}{c}{\textbf{MIS\_micro}} \\
\cmidrule(lr){2-7} \cmidrule(lr){8-13} \cmidrule(lr){14-19}
\textbf{Model}
& \multicolumn{3}{c}{\textbf{BigO}} & \multicolumn{3}{c}{\textbf{CC}}
& \multicolumn{3}{c}{\textbf{BigO}} & \multicolumn{3}{c}{\textbf{CC}}
& \multicolumn{3}{c}{\textbf{BigO}} & \multicolumn{3}{c}{\textbf{CC}} \\
\cmidrule(lr){2-4} \cmidrule(lr){5-7} \cmidrule(lr){8-10} \cmidrule(lr){11-13} \cmidrule(lr){14-16} \cmidrule(lr){17-19}
& 0 & 0.7 & 0.95 & 0 & 0.7 & 0.95 & 0 & 0.7 & 0.95 & 0 & 0.7 & 0.95 & 0 & 0.7 & 0.95 & 0 & 0.7 & 0.95 \\
\midrule

% --- Commercial Models ---
\multicolumn{19}{@{}l}{\cellcolor{gray!20}\textbf{Commercial Models}} \\
\textbf{\textit{Language/Code Models}} \\
GPT-4o
& 0.68 & 0.73 & 0.80 & 0.15 & 0.17 & 0.19
& 0.0042 & 0.0058 & 0.0072 & 0.0097 & 0.0091 & 0.0130
& 0.0068 & 0.0110 & 0.0132 & 0.0097 & 0.0091 & 0.0130 \\
Claude-3.7-Sonnet
& 0.77 & 0.81 & 0.87 & 0.22 & 0.23 & 0.25
& 0.0026 & 0.0029 & 0.0032 & 0.0113 & 0.0107 & 0.0100
& 0.0023 & 0.0028 & 0.0030 & 0.0113 & 0.0107 & 0.0100 \\
gpt-3.5-turbo-instruct
& 0.35 & 0.38 & 0.43 & - & - & -
& 0.0023 & -- & -- & 0.0010 & 0.0043 & 0.0050
& 0.0024 & -- & -- & 0.0020 & 0.0072 & 0.0089 \\
\addlinespace
\textbf{\textit{Reasoning Model}} \\
GPT-o4-mini
& 0.69 & 0.72 & 0.80 & 0.11 & 0.11 & 0.12
& 0.0029 & 0.0109 & 0.0126 & 0.0096 & 0.0113 & 0.0115
& 0.0084 & 0.0092 & 0.0098 & 0.0096 & 0.0113 & 0.0115 \\
\midrule

% --- Open Source Models ---
\multicolumn{19}{@{}l}{\cellcolor{gray!20}\textbf{Open Source Models}} \\
\textbf{\textit{Code Models}} \\
Qwen2.5-Coder-7B
& 0.39 & 0.41 & 0.46 & 0.05 & 0.05 & 0.06
& 0.0000 & 0.0079 & 0.0096 & 0.0050 & 0.0004 & 0.0005
& 0.0076 & 0.0098 & 0.0119 & 0.0050 & 0.0004 & 0.0005 \\
CodeLlama-7B-Instruct
& 0.10 & 0.11 & 0.12 & 0.01 & 0.01 & 0.01
& 0.0033 & 0.0043 & 0.0067 & 0.0095 & 0.0067 & 0.0139
& 0.0033 & 0.0049 & 0.0058 & 0.0095 & 0.0067 & 0.0139 \\
Codestral-22B
& 0.48 & 0.50 & 0.55 & 0.09 & 0.10 & 0.11
& 0.0025 & 0.0029 & 0.0031 & 0.0072 & 0.0060 & 0.0072
& 0.0025 & 0.0029 & 0.0032 & 0.0072 & 0.0060 & 0.0072 \\
\addlinespace
\textbf{\textit{Language/Code Models}} \\
Llama-3.1-8B
& 0.28 & 0.30 & 0.34 & 0.06 & 0.06 & 0.07
& 0.0030 & 0.0041 & 0.0044 & 0.0015 & 0.0020 & 0.0022
& 0.0030 & 0.0036 & 0.0038 & 0.0015 & 0.0020 & 0.0022 \\
Mistral-7B-v0.3
& 0.10 & 0.11 & 0.12 & 0.01 & 0.01 & 0.01
& 0.0028 & 0.0033 & 0.0049 & 0.0080 & 0.0052 & 0.0102
& 0.0029 & 0.0036 & 0.0044 & 0.0080 & 0.0052 & 0.0102 \\
\addlinespace
\textbf{\textit{Reasoning Models}} \\
DeepSeek-R1-Distill-Llama-70B
& 0.13 & 0.14 & 0.15 & 0.01 & 0.01 & 0.01
& 0.0026 & 0.0029 & 0.0034 & 0.0075 & 0.0045 & 0.0071
& 0.0027 & 0.0031 & 0.0034 & 0.0075 & 0.0045 & 0.0071 \\
DeepSeek-R1-Distill-Qwen-32B
& 0.15 & 0.15 & 0.16 & 0.01 & 0.01 & 0.01
& 0.0029 & 0.0034 & 0.0038 & 0.0083 & 0.0053 & 0.0079
& 0.0030 & 0.0036 & 0.0038 & 0.0083 & 0.0053 & 0.0079 \\
\bottomrule
\end{tabular}
} 
\end{table*}

\begin{table*}[t]
\centering
\caption{Sensitivity of instability metrics to sampling knobs (\textbf{BigOBench}) for \textbf{GPT-o4-mini} at \textbf{temp = 0.7}}
\label{tab:hyperparam_sensitivity}
\scriptsize
\setlength{\tabcolsep}{4pt}
\renewcommand{\arraystretch}{0.95}
\begin{tabular}{@{} l c c c c c c c c c @{}}
\toprule
\textbf{Setting} & \textbf{Mode} & \textbf{Stride $s$} & \textbf{Quant. $q$ (B)} & \textbf{Interval $\Delta t$ (ms)} &
\textbf{MIS\_macro} & \textbf{$\Delta$macro \%} &
\textbf{MIS\_micro} & \textbf{$\Delta$micro \%} &
\textbf{Paired $\tilde{\Delta}D_p$ [IQR] \quad $p$ / $\delta$} \\
\midrule
\rowcolor{gray!15}
\textbf{Baseline} & Line & 1 & 64 & -- &
\textbf{0.0045} & -- &
\textbf{0.0065} & -- &
-- \\
\addlinespace[1pt]
Stride $\uparrow$ & Line & 5 & 64 & -- &
0.0047 & +5\% &
0.0070 & +8\% &
$+0.00015$ [\;0.0018\;] \quad $p{=}0.06$ / $\delta{=}0.16$ \\
 & Line & 10 & 64 & -- &
0.0048 & +6\% &
0.0071 & +9\% &
$+0.00018$ [\;0.0018\;] \quad $p{=}0.05$ / $\delta{=}0.18$ \\
\addlinespace[2pt]
Quantization $\uparrow$ & Line & 1 & 128 & -- &
0.0045 & +0.5\% &
0.0066 & +0.8\% &
$+0.00002$ [\;0.0002\;] \quad $p{=}0.70$ / $\delta{=}0.02$ \\
 & Line & 1 & 256 & -- &
0.0045 & +0.8\% &
0.0066 & +1.0\% &
$+0.00003$ [\;0.0002\;] \quad $p{=}0.62$ / $\delta{=}0.03$ \\
\addlinespace[2pt]
Time sampling & Time & -- & 64 & 0.1 &
0.0066 & +46\% &
0.0101 & +55\% &
$+0.00160$ [\;0.0018\;] \quad $p{<}0.001$ / $\delta{=}0.62$ \\
(matched density) & Time & -- & 64 & 1 &
0.0053 & +18\% &
0.0080 & +23\% &
$+0.00090$ [\;0.0012\;] \quad $p{=}0.010$ / $\delta{=}0.36$ \\
 & Time & -- & 64 & 2 &
0.0050 & +10\% &
0.0074 & +14\% &
$+0.00055$ [\;0.0008\;] \quad $p{=}0.07$ / $\delta{=}0.22$ \\
\bottomrule
\end{tabular}

\vspace{2pt}
\footnotesize
\textbf{Notes}: Paired $\tilde{\Delta}D_p$ = median across problems of $D_p^{\text{setting}}-D_p^{\text{baseline}}$; IQR = interquartile range of these differences; $p$ = Wilcoxon signed-rank $p$-value; $\delta$ = Cliff’s delta.
\end{table*}

\begin{table*}
\centering
\scriptsize
\caption{Stability proxies vs.\ SE metrics}
\label{tab:se_from_stability_one_table}
\resizebox{\textwidth}{!}{
\begin{tabular}{@{}lcccccc@{}}
\toprule
& \multicolumn{2}{c}{\textbf{Cognitive Complexity (mean)}} 
& \multicolumn{2}{c}{\textbf{Maintainability Index}} 
& \multicolumn{2}{c}{\textbf{Cyclomatic Complexity (mean)}} \\
\cmidrule(lr){2-3} \cmidrule(lr){4-5} \cmidrule(lr){6-7}
\textbf{Model} 
& \textbf{DMPD} ($\rho\,|\,\delta$) & \textbf{NMV} ($\rho\,|\,\delta$)
& \textbf{DMPD} ($\rho\,|\,\delta$) & \textbf{NMV} ($\rho\,|\,\delta$)
& \textbf{DMPD} ($\rho\,|\,\delta$) & \textbf{NMV} ($\rho\,|\,\delta$) \\
\midrule
Claude-3.7-Sonnet
& $-0.094 \,|\, 0.121$ & $0.334 \,|\, -0.405$
& $0.180 \,|\, -0.260$ & $0.050 \,|\, -0.125$
& $-0.134 \,|\, 0.187$ & $0.255 \,|\, -0.288$ \\
Codestral-22B
& $0.098 \,|\, -0.055$ & $0.413 \,|\, -0.497$
& $0.187 \,|\, -0.296$ & $0.035 \,|\, -0.165$
& $0.071 \,|\, -0.016$ & $0.346 \,|\, -0.401$ \\
DeepSeek-R1-Distill-Llama-70B
& $0.316 \,|\, -0.377$ & $0.545 \,|\, -0.671$
& $0.752 \,|\, -0.889$ & $0.362 \,|\, -0.502$
& $0.024 \,|\, -0.066$ & $0.455 \,|\, -0.571$ \\
DeepSeek-R1-Distill-Qwen-32B
& $0.433 \,|\, -0.388$ & $0.767 \,|\, -1.000$
& $0.128 \,|\, -0.265$ & $0.232 \,|\, -0.297$
& $0.310 \,|\, -0.250$ & $0.697 \,|\, -0.918$ \\
Llama-3.1-8B
& $-0.178 \,|\, 0.110$ & $0.471 \,|\, -0.550$
& $0.354 \,|\, -0.461$ & $0.528 \,|\, -0.707$
& $-0.288 \,|\, 0.304$ & $0.429 \,|\, -0.465$ \\
\bottomrule
\end{tabular}
}
\vspace{2pt}
{\footnotesize
\textbf{Reading guide:} Tertiles are computed \emph{per model} and \emph{separately} for \textsc{DMPD} (instability) and \textsc{NMV} (burstiness): T1{=}low proxy (more stable), T3{=}high proxy (less stable). Cells report Spearman’s $\rho$ and Cliff’s $\delta(\text{T1},\text{T3})$; since T3 is less stable, $\delta<0$ means the SE metric tends to be \emph{higher} when instability/burstiness is higher.
}
\end{table*}

\begin{figure*}[t]
  \centering
  \vspace{-1em}
  \includegraphics[page=1,width=0.65\textwidth]{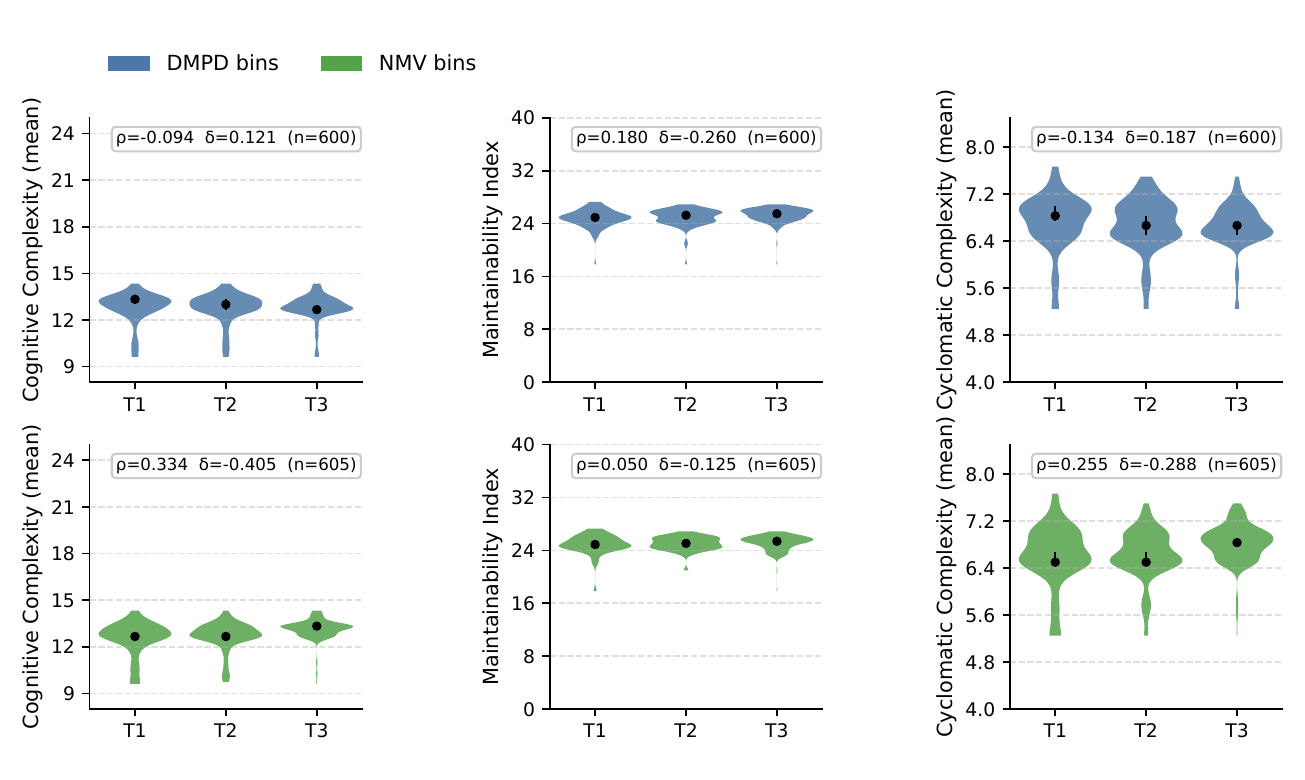}
  \vspace{-2em}
  \caption{Claude-3.7-Sonnet — Stability tertiles (T1 low → T3 high) vs SE metrics}
  \label{fig:violen_plot_with_SE}
  \vspace{-1em}
\end{figure*}

\subsection{Relationship Between Operational Stability and Code Quality}
\label{sec: to_se}
As discussed before higher (\textsc{DMPD}) values indicate \emph{greater instability},
now higher (\textsc{NMV}) values can be thought of as \emph{greater burstiness}. We form tertiles on the \emph{raw} proxy values separately
for \textsc{DMPD} and \textsc{NMV} (T1{=}lowest, T3{=}highest). Hence, ``moving from T1 to T3''
means \emph{decreasing} stability (for \textsc{DMPD}) or \emph{increasing} burstiness (for \textsc{NMV}).
Cliff’s $\delta$ is reported as $\delta(\text{T1},\text{T3})$; thus $\delta<0$ means the SE metric is
\emph{larger at higher instability/burstiness}. The double-violin plot in Figure~\ref{fig:violen_plot_with_SE} shows the spread and relationship for Claude. 

Across models, we observe two robust regularities under \textsc{NMV} tertiles:
(i) \emph{Cognitive} and \emph{Cyclomatic} complexities consistently increase as burstiness rises
($\delta<0$ in all reported models; $\rho>0$), and
(ii) \emph{Maintainability Index} (MI) also tends to increase with burstiness
(negative $\delta$ with small-to-moderate positive $\rho$)(Refer to Table~\ref{tab:se_from_stability_one_table}).
Under \textsc{DMPD} tertiles, the most stable pattern concerns MI: higher \textsc{DMPD} is associated with \emph{higher} MI, whereas the relations to complexity
metrics are comparatively weak and model-dependent (signs of $\delta$ are mixed and magnitudes
are smaller). Taken together, these results indicate that the two proxies emphasize different
facets of code behavior: \textsc{NMV} (burstiness) aligns strongly with structural complexity,
while \textsc{DMPD} aligns more with MI.
\begin{paperbox}[Answer to RQ5]{analysisbg}
\textbf{Instability is moderately linked to\ SE quality.} Higher \textsc{NMV} (burstiness) tracks \emph{higher} Cognitive/Cyclomatic complexity, while higher \textsc{DMPD} (higher instability) aligns mainly with \emph{higher} MI; links to complexity under \textsc{DMPD} are weak/inconsistent.
\end{paperbox}
\vspace{-1em}

%% file: sections/discussion.tex
\section{Discussion}

\subsection{Cliffs of Correctness}
LLM-authored code often exhibits \emph{correctness cliffs}: under a fixed prompt, tiny perturbations like temperature or random seed can flip a sample from fail to pass, yielding sharp jumps in pass@\({k}\) as \(k\) increases \citep{chen2021evaluating}. The same stochastic variability extends beyond correctness to \emph{operational shape}. Two candidates that both pass unit tests may realize markedly different allocation patterns (temporary buffers, cache lifetimes, data structure choices). At small \(N\) these differences look benign, yet modest input or traffic shifts can push one variant over a tail threshold: an OOM kill, paging storm, or latency blowup, while the other remains stable. Classic systems guidance reminds us that user and business impact are dominated by \emph{tails}, not means \citep{dean2013tail}. 

\textit{Use in post-training.}
Ideally, the manifold of \emph{correct} solutions should be \emph{flat}, meaning nearby samples behave similarly at runtime rather than riddled with cliffs. Although injecting execution signals directly into pretraining is costly and often non-differentiable, they can be exploited in post-training (e.g., policy gradient methods like PPO \cite{schulman2017proximal}) to smooth execution effects by rewarding low \textsc{DMPD} (down-weighting unstable candidates) in post-training. We do not claim to cover all operational risks. Instead, we report associations between runtime-stability proxies (\textsc{DMPD}/\textsc{MIS}) and established SE maintainability metrics, and leave causal links to production SLOs as future work.
\vspace{-1em}

\subsection{Implications for Auto-Generated Software}
\label{sec:implications_auto_generated_software}
Our results show that instability affects \emph{software qualities} even when functional tests pass. As burstiness rises (\textsc{NMV}: T1$\!\to$T3), we consistently observe \emph{higher} Cognitive and Cyclomatic complexity, while Maintainability Index (MI) also trends upward. Additionally, lower stability by \textsc{DMPD} is most reliably associated with \emph{higher} MI, with weak links to cognitive or cyclomatic complexity. 

\paragraph{Why this relation matters in practice.}
In a typical pipeline, teams gate on "working" solutions, shipping whichever sample happens to pass the tests. Yet two passing candidates can embody very different \emph{runtime stories}: one grows memory smoothly; another idles for a while and then surges late. Nothing breaks at small inputs, so the difference might be invisible until a real-life load situation nudges the spikier variant over a limit. The result is not just an isolated OOM or retry cascade; it is a stream of hidden costs that accumulate over time—extra headroom in capacity planning, flaky builds, noisy alerts, and engineering toil. These effects compound across services and releases, creating operational drag that correctness metrics simply do not surface \cite{beyer2016site, nygard2018release, dean2013tail}.

\paragraph{Practical selection.}
Because these costs add up, it is worth choosing not just a \emph{working} candidate, but the \emph{most stable} one and, where possible, the model that is most stable for the task at hand. A lightweight policy is enough: for each patch, sample \(k\) correct generations, run a brief single-bin scale check, compute \textsc{DMPD}/\textsc{NMV}, and select the candidate with low median and tight spread (subject to correctness and basic latency checks). At the model level, prefer those with lower \(\mathrm{MIS}\) on your workload. This adds minimal overhead to CI but directly reduces the risk that small, everyday changes snowball into production incidents \cite{humble2010continuous, beyer2018site}.

\subsection{Why DMPD robustness matters in industry.}

DMPD offers a version-aware signal that is comparatively less confounded by the \emph{scale} and \emph{composition} of the test workload than peak or normalized peak memory. Indeed, in large repositories, peak memory can fluctuate primarily because the workload grows, the test suite evolves, or integration coverage shifts. This makes peak-based comparisons fragile for pre vs post patch assessment. By emphasizing the \emph{shape} change of memory-profile behavior between versions, DMPD provides a more test condition-agnostic view of whether a patch meaningfully alters execution dynamics, which is precisely the sort of comparative evidence that teams seek when reasoning about operational risk before release.

%% file: sections/threats_to_validity.tex
\vspace{-1em}
\section{Threats to Validity}

\paragraph{Language and runtime specificity.}
All experiments use Python and instrument application-level allocations via \texttt{tracemalloc} under CPython’s reference counting plus cyclic GC \cite{pep454,gcdocs}. These choices suppress RSS/allocator noise but also bias us toward Python’s memory semantics. Other ecosystems differ materially e.g., JVM and .NET employ moving, generational collectors with tunable heuristics \cite{jayasena2015auto,jones2023garbage}, Go prioritizes low-latency concurrent GC \cite{hudson2015go}, while Rust enforces ownership and largely avoids GC \cite{klabnik2023rust}. Such differences can change allocation shapes and, therefore, DTW distances. Low-level allocator behavior and fragmentation also vary by platform \cite{valgrind2007,jones2023garbage}. Hence, our scores cannot be transferred to other languages without replication.

\paragraph{Benchmark vs.\ real systems.}
We evaluate on simple competitive coding style problems (CodeContests, BigOBench), not on production services. Real software is more complex and introduces frameworks, concurrency, and deployment constraints (containers, limits, autoscaling) that may interact with memory trajectories and tail behavior \cite{dean2013tail}. External validity is a general challenge in SE empiricism \cite{wohlin2012m, kitchenham2007guidelines}. Our findings motivate, but do not replace, validation on industrial repositories and end-to-end workloads.

\paragraph{Sampling budget and statistical power.}
We cap generations to each problem at \(N{=}5\) correct generations and at most \(r{=}10\) private tests. This modest budget limits \(\binom{k}{2}\) pairwise comparisons, can miss long-tail behaviors, and increases the variance of \(D_p\) and \(\mathrm{MIS}\). Although we fix seeds and use a single-bin policy to improve repeatability (for Mercury scaling test), estimates remain sensitive to which samples happen to be drawn. We report medians/IQRs and perform sensitivity checks, but larger \(N\)/\(r\) or explicit uncertainty estimates would yield tighter measurements.

%% file: sections/relwork.tex
\vspace{-1em}
\section{Related Work}
Research on code–generating LLMs has moved from “does it work?” to “how well does it behave?”, with a growing emphasis on properties beyond pure I/O correctness. We outline this trajectory below.

\paragraph{Phase I: Functional correctness at scale.}
Early benchmarks established that LLMs can produce \emph{passing} programs, measuring \texttt{pass@k} on HumanEval and MBPP \cite{chen2021evaluating, austin2021program}. Large-scale competitive settings (e.g., AlphaCode/CodeContests) operationalized this lens by filtering massive candidate pools through public/private tests \cite{li2022competition}. These efforts treat all passing solutions as equivalent and do not distinguish \emph{how} they behave at runtime.

\paragraph{Phase II: Text and structure beyond I/O.}
To differentiate among multiple passing outputs, evaluation broadened to surface form and program structure. CodeBLEU augments n-gram matching with syntax. \cite{ren2020codebleu}. Structural similarity metrics: AST-based measures and tree-edit distances such as TSED quantify how two programs differ in static structure\cite{jiang2007deckard,song2024revisiting}. Empirical audits further showed that LLM code, while correct, can be brittle or suboptimal \cite{tian2023chatgpt}. These approaches, however, remain \emph{representation-level} proxies; they do not look at dynamic behavior under execution.

\paragraph{Phase III: Execution-grounded signals.}
Recent work exploits runtime signals directly. AlphaCode leverages test outcomes to triage candidates \cite{li2022competition}; AlphaDev optimizes measured latency to discover faster algorithms \cite{mankowitz2023faster}; energy-aware generation explores efficiency objectives \cite{ilager2025green}. In parallel, prompting for algorithmic complexity (\textsc{BigOBench}) connects predicted complexity to time/space behavior \cite{chambon2025bigo}. Outside the LLM literature, mature profilers (e.g., Python \texttt{tracemalloc}/PEP~454 and Valgrind/Massif) provide the instrumentation needed to observe allocation dynamics \cite{pep454,valgrind2007}. Despite these ingredients, very few systematically quantify the stability of runtime memory behavior across multiple correct generations of the same specification.

%% file: sections/conclusion.tex
\section{Conclusion}
Correctness isn’t consistency: even \emph{passing} LLM programs can diverge in runtime memory behavior. We introduced MPP (to denoise traces) and a shape-based DMPD aggregated as MIS (with NMV for burstiness) to quantify this instability. Beyond being a measurement gap, instability imposes \emph{hidden software-engineering costs} extra capacity headroom, flaky builds, noisy alerts, incident toil, and long-run maintainability drag—that pass@k and text/syntax metrics do not surface. Practically, teams should pick the \emph{stable} passing candidate (low DMPD/NMV) and prefer models with lower MIS, e.g., via stability-aware reranking and conservative temperatures. Looking forward, extending these ideas beyond Python/memory to latency/CPU/I/O and linking stability to SLOs and incident data will better translate runtime shape into operational risk.